\documentclass[10pt]{iopart}

\usepackage{iopams}
\usepackage[english]{babel}
\usepackage{graphicx}
\usepackage{cite}
\usepackage{color}
\usepackage{bbm}

\begin{document}

% Full title of the paper (Capitalized)
\title{Revisiting the optical $\mathcal{PT}$-symmetric dimer}

% Authors, for the paper (add full first names)
\author{J. D. Huerta Morales}
\address{Instituto Nacional de Astrof\'{\i}sica, \'Optica y Electr\'onica, Calle Luis Enrique Erro No. 1, Sta. Ma. Tonantzintla, Pue. CP 72840, M\'exico}
\ead{jd\_huerta@inaoep.mx}

\author{J. Guerrero}
\address{Departamento de Matem\'atica Aplicada, Facultad de Inform\'atica, Campus Espinardo, Univesidad de Murcia, 30100 Murcia, Spain.}
\ead{juguerre@um.es}

\author{S. L\'opez-Aguayo}
\address{Photonics and Mathematical Optics Group, Tecnol\'ogico de Monterrey, Monterrey 64849, Mexico.}
\ead{servando@itesm.mx}

\author{B. M. Rodr\'iguez-Lara}
\address{Instituto Nacional de Astrof\'{\i}sica, \'Optica y Electr\'onica, Calle Luis Enrique Erro No. 1, Sta. Ma. Tonantzintla, Pue. CP 72840, M\'exico}
\ead{bmlara@inaoep.mx}

\begin{abstract}
Optics has proved a fertile ground for the experimental simulation of quantum mechanics.
Most recently, optical realizations of $\mathcal{PT}$-symmetric quantum mechanics have been shown, both theoretically and experimentally, opening the door to international efforts aiming at the design of practical optical devices exploiting this symmetry.
Here, we focus on the optical $\mathcal{PT}$-symmetric dimer, a two-waveguide coupler were the materials show symmetric effective gain and loss, and provide a review of the linear and nonlinear optical realizations from a symmetry based point of view.
We go beyond a simple review of the literature and show that the dimer is just the smallest of a class of planar $N$-waveguide couplers that are the optical realization of Lorentz group in 2+1 dimensions.
Furthermore, we provide a formulation to describe light propagation through waveguide couplers described by non-Hermitian mode coupling matrices based on a non-Hermitian generalization of Ehrenfest theorem.
\end{abstract}

\maketitle

\section{Introduction}

The desire to create an optical directional coupler, a device composed by parallel optical waveguides close enough that leaked energy is transferred between them, led to the exploration of waveguide creation in semi-conductors via proton bombardment \cite{Garmire1972p87}.
At the time, power losses played an interesting role and the nascent mode coupling theory \cite{Marcatili1969p2071} allowed the theoretical description of linear loses in such devices \cite{Somekh1973p46},
\begin{eqnarray}\label{eq:SomekhCoupler}
-i \frac{d}{dz} E_{n}(z) = \frac{i \alpha}{2} E_{n}(z) + K \left[ E_{n-1}(z) + E_{n+1}(z) \right],
\end{eqnarray}
where the real numbers $\alpha$ and $K$ are the effective linear loss, identical in all implanted waveguides, and the effective waveguide coupling strength, also indentical for the whole system, in that order.
This, to the best of our knowledge, was the first theoretical description of an experimental $N$-waveguide coupler including losses in the form of a Schr\"odinger-like equation involving a non-Hermitian Hamiltonian.
Almost twenty years later, the desire to create an intensity dependent switch working at low power levels took another team of researchers to explore twin core nonlinear couplers with gain and loss.
Again, coupled mode theory allowed the description of such devices \cite{Chen1992p239},
\begin{eqnarray}
-i \frac{d}{dz} E_{1}(z) &=& \left( \beta_{1} + i \alpha_{1}  \right) \mathcal{E}_{1} + \Delta\beta_{1}\left( \vert \mathcal{E}_{1} \vert^{2}\right) \mathcal{E}_{1} + K \mathcal{E}_{2}, \\
-i \frac{d}{dz} E_{2}(z) &=& \left( \beta_{2} + i \alpha_{2}  \right) \mathcal{E}_{2} + \Delta\beta_{2}\left( \vert \mathcal{E}_{2} \vert^{2}\right) \mathcal{E}_{2}+ K \mathcal{E}_{1},
\end{eqnarray}
where we have kept the notation used before and introduce the effective real part of the refractive index, $\beta_{j}$, and the real function $\Delta\beta_{j}\left( \vert \mathcal{E}_{j} \vert^{2}\right) = \pm \kappa_{j} \vert \mathcal{E}_{j} \vert^{2} $ that describes an effective Kerr nonlinearity induced change in the refractive index of the $j$th core, positive for self-focusing and negative for self-defocusing materials.
At this point in history, there existed experimental and theoretical work describing an optical dimer where the waveguides present effective loss and gain and a nonlinearity but a little something was missing.
A couple of years later, a theory exploring a particular type of non-Hermitian Hamiltonians with real spectra was brought forward in quantum mechanics \cite{Bender1998p5243}.
These Hamiltonians were invariant under space-time reflection, received the name of $\mathcal{PT}$-symmetric, and opened new avenues of research in quantum mechanics as well as other areas of physics and matemathics, c.f. \cite{Bender2005p277,Bender2007p947} and references therein.

In optics, it took a few years more to propose two seminal ideas. The first one regarded single elements and showed, in particular, that an optical planar slab waveguide composed of two media with linear gain and loss can be described by a Schr\"odinger-like equation under dynamics dictated by a $PT$-symmetric Hamiltonian, where the optical refractive index played the role of quantum-like potential and propagation distance that of time \cite{Ruschhaupt2005p171}.
The second one dealt with composite systems, where a mature mode coupling theory produced a theory of coupled optical PT-symmetric structures \cite{ElGanainy2007p2632}.
In the ten years following those first proposals for an optical realization of $PT$-symmetry, work has been reported on slab waveguides \cite{Klaiman2008p080402,Mostafazadeh2009p220402}, Bragg scatterers \cite{Berry2008p244007,Longhi2010p031801,Longhi2011p042119,Lin2011p213901,Miri2013p043819,Kozlov2015p105004}, as well as linear \cite{Makris2008p103904,Longhi2009p123601,Vemuri2011p043826,Ramezani2012p013818,Vemuri2013p044101,Joglekar2013p30001,Longhi2014p1697,Chern2015p5806} and nonlinear \cite{Musslimani2008p030402,Ramezani2010p043803,Sukhorukov2010p043818,Regensburger2012p167,Sukhorukov2012p2148,Kevrekidis2013p365201,Barashenkov2013p053817,Khawaja2013p023830,Lumer2013p263901,Barashenkov2014p045802,Zhang2014p13927,Barashenkov2015p325201,Martinez2015p023822,Walasik2015p19826,Cole2016p013803,Kartashov2016p013841} coupled waveguides, to mention just a few.
Research in this field is slowly getting to information technologies applications with recent proposals of all-optical $\mathcal{PT}$-symmetric logic gates \cite{Ding2015p123104} and amplitude-to-phase converters \cite{ZaragozaGutierrez2015p3989}.

Here, we will provide a review of the optical $\mathcal{PT}$-symmetric dimer.
First, in Section 2, we will introduce a two-waveguide coupler where component waveguides show effective complex refractive indices with identical real part.
Starting from this device, we will recover the effective mode coupling differential equation set for the linear $\mathcal{PT}$-symmetric dimer, which describes a nonunitary optical device showing symmetric effective loss and gain feasible of passive and active optical realizations.
Then, we will recover the dispersion relation for the dimer that shows three regimes, one with real eigenvalues, the $\mathcal{PT}$-symmetric regime, another with fully degenerate eigenvalues equal to zero, the fully degenerate regime, and a third one with purely imaginary eigenvalues, the broken symmetry regime.
We will construct an analytic propagator that will show asymmetric amplifying oscillator, power amplification, and exponential amplification behaviors in each of these regimes.
We will also show that it is possible to uncouple the mode coupling differential equation set of the $\mathcal{PT}$-symmetric dimer.
The resulting second order differential equations and boundary conditions for the field amplitudes propagating through each waveguides take the form of nonlocal oscillators with positive potential, free particle traveling through a nonlocal medium, and nonlocal oscillators with inverted potential, in each of the regimes.
In the final part of Section 2, we will bring forward the renormalized field approach that helps us cast the linear $\mathcal{PT}$-symmetric dimer as a nonlinear dimer with imaginary Kerr nonlinearity, either in a self- or a cross-modulation scheme, and allows us to realize an asymptotic behavior that depends just on the gain to coupling strength ratio of the device.
In section 3, we will discuss the linear $\mathcal{PT}$-symmetric dimer when both waveguides show the same effective self-focusing Kerr nonlinearity.
We will show the stable nonlinear modes of the device, discuss its dynamics in terms of the passive Kerr two-waveguide coupler that allows for coherent and localized oscillations between the waveguide field modes, and show that the inclusion of symmetric gain and loss breaks these dynamics producing localization in the gain waveguide without showing an asymptotic behavior.
In Section 4, we will extend the linear $\mathcal{PT}$-symmetric dimer to planar $N$-waveguide couplers using finite dimensional matrix representations of a complexified version of $SU(2)$.
We have previously shown \cite{RodriguezLara2015p5682} that the $\mathcal{PT}$-symmetric dimer and its extensions to planar $N$-waveguide couplers possess an $SO(2,1)$ symmetry realized in a finite dimensional non-unitary irreducible representation.
This representation is accomplished through complexification of $SU(2)$, $\{ \mathbbm{J}_{x}, \mathbbm{J}_{y}, \mathbbm{J}_{z}\} \rightarrow \{ \mathbbm{J}_{x}, i \mathbbm{J}_{y}, i \mathbbm{J}_{z}\}\equiv \{\mathbbm{K}_{z},\mathbbm{K}_{x},\mathbbm{K}_{y}\}$, and it allows us to provide the dispersion relation and a closed form analytic propagator, which have the same regimes and dynamics found for the dimer.
We will show that the renormalized field approach provides us with an asymptotic behavior that is independent of the initial field distribution and depends just on the waveguide number and the effective gain to coupling ratio.
Then, in section 5, we will introduce a modified version of the Ehrenfest theorem suitable for non-Hermitian Hamiltonians and show how it can help us define the dynamics of an $N^{2}$-dimensional generalized Stokes vector for the planar $N$-waveguide couplers discussed in section 4.
In section 6, we will go back to the dimer but consider the propagation of quantum fields.
In the quantum regime, spontaneous generation and absorption of electromagnetic radiation should be considered when using media with linear gain or loss.
We will show the solution for the quantum linear $\mathcal{PT}$-symmetric dimer and discuss the generation of light from vacuum due to spontaneous processes in the absence of fields impinging at the device.
Finally, we will produce a brief summary and discuss future avenues regarding non-Hermitian optical systems.

%%%%%%%%%%%%%%%%%%%%%%%%%%%%%%%%%%%%%%%%%%
\section{Linear $\mathcal{PT}$-symmetric dimer}\label{Sec:LinearPTdimer}

A two-waveguide coupler can be described by the following differential equation system via mode coupling theory,
\begin{eqnarray}
- i \partial_{z} \left( \begin{array}{c} E_{1}(z) \\ E_{2}(z) \end{array}\right) = \left( \begin{array}{cc}
n_{1} & g \\ g & n_{2}  \end{array} \right) \left( \begin{array}{c} E_{1}(z) \\ E_{2}(z) \end{array}\right),
\end{eqnarray}
where the complex numbers $E_{j}$ are the field amplitudes at each waveguide, the parameters $n_{j}$ and $g$, which in the most general case can be complex, are the effective refractive indices and  waveguide coupling, in that order, and the operator $\partial_{z}$ stands for the derivative with respect to the propagation distance $z$.
Note that using field amplitudes of the form
\begin{equation}
E_{j}(z) = e^{  i n_{+} z } \mathcal{E}_{j}(z),
\end{equation}
with $n_{\pm}= \left( n_{1} \pm n_{2} \right)/2$, reduces the system to an effective mode coupling matrix,
\begin{eqnarray}
- i \partial_{z} \left( \begin{array}{c} \mathcal{E}_{1}(z) \\ \mathcal{E}_{2}(z) \end{array}\right) = \left( \begin{array}{cc}
n_{-} & g \\ g & -n_{-}  \end{array} \right) \left( \begin{array}{c} \mathcal{E}_{1}(z) \\ \mathcal{E}_{2}(z) \end{array}\right), \label{eq:PTDimerMC}
\end{eqnarray}
The mode coupling matrix has eigenvalues
\begin{eqnarray}
\varepsilon_{\pm} = \pm \sqrt{g^{2} +   n_{-}^{2}}.
\end{eqnarray}
The eigenvalues are real for the case of identical real part of the effective refractive indices, $\Re (n_{-}) = 0$, and imaginary part less than the value of the effective coupling, $\Im (n_{-}) < g $.
The eigenvalues degenerate in the case $\Re (n_{-}) = 0$ and $\Im (n_{-}) = g$.
They are purely imaginary for $\Re( n_{-} ) = 0$ and $\Im ( n_{-} ) > g$  and, finally, complex elsewhere.
This general non-Hermitian dimer has a rich structure that deserves further attention but, right now, we are interested in just the $\mathcal{PT}$-symmetric case.

\begin{figure}[h]
\centering
\includegraphics[width=\textwidth]{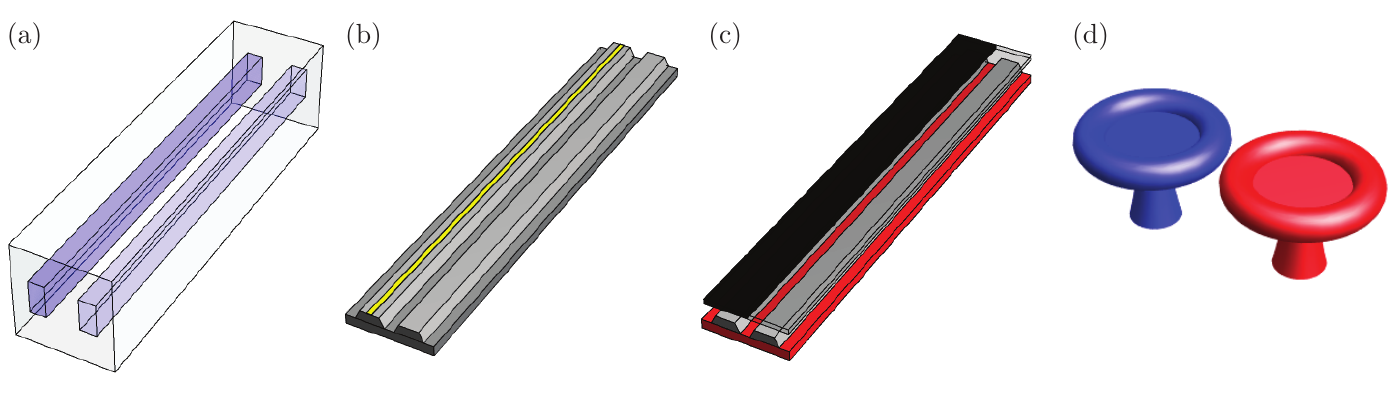}
\caption{Schematics of experimental realizations of the linear $\mathcal{PT}$-symmetric dimer. (a) Passive laser engraved waveguides, (b) passive waveguides with metallic scatterers, (c) pumped active waveguides, (d) pumped active whispering-gallery mode microcavities.}
\label{fig:Figure1}
\end{figure}

In quantum mechanics, PT-symmetry refers to space-time reflection symmetry \cite{Bender2005p277}.
In discrete optical couplers we can consider waveguide permutation and propagation inversion as equivalent to space and time reflection, respectively.
Then, in order to recover the standard linear $PT$-symmetric dimer, we need to work with waveguides that have the same effective refractive indices, $\Re( n_{1} )= \Re(n_{2})$ such that $n_{-}$ is purely imaginary and we can write a differential set \cite{RodriguezLara2015p5682},
\begin{eqnarray} \label{eq:PTDimer}
- i \partial_{\zeta} \left( \begin{array}{c} \mathcal{E}_{1}(\zeta) \\ \mathcal{E}_{2}(\zeta) \end{array}\right) = \left( \begin{array}{cc}
i \gamma & 1 \\ 1 & -i \gamma  \end{array} \right) \left( \begin{array}{c} \mathcal{E}_{1}(\zeta) \\ \mathcal{E}_{2}(\zeta) \end{array}\right),
\end{eqnarray}
where permutation of the waveguides, $\pm i \gamma \rightarrow \mp i \gamma$, and propagation reversal, $\zeta \rightarrow -\zeta$, leaves the system invariant.
Note that we have used the effective coupling parameter to  scale the propagation distance, $\zeta = g z$, such that we deal with a single parameter given by the effective refractive index to coupling ratio, $\gamma = \Im ( n_{-} ) / g$ with $\Re(n_{-}) = 0$.
This mathematical model is equivalent to consider waveguides with effective pure linear loss and gain, $i \gamma$ and $-i \gamma$ in that order.
In optics, the linear $\mathcal{PT}$-symmetric dimer has been experimentally demonstrated in passive lossy waveguides \cite{Guo2009p093902,Ornigotti2014p065501}, Fig. \ref{fig:Figure1}(a) and Fig. \ref{fig:Figure1}(b), as well as active, pumped waveguides \cite{Ruter2010p192}, Fig. \ref{fig:Figure1}(c), and pumped whispering-gallery mode microcavities \cite{Peng2014p394,Hodaei2014p975}, Fig. \ref{fig:Figure1}(d), with linear gain.
The experimental demonstration of $\mathcal{PT}$-symmetric devices is not limited to optical resonators, they have also been realized with operational amplifiers in electronics \cite{Schindler2012p444029}.

\subsection{Quantum mechanics, linear algebra approach}
\begin{figure}[h]
\centering
\includegraphics[width=\textwidth]{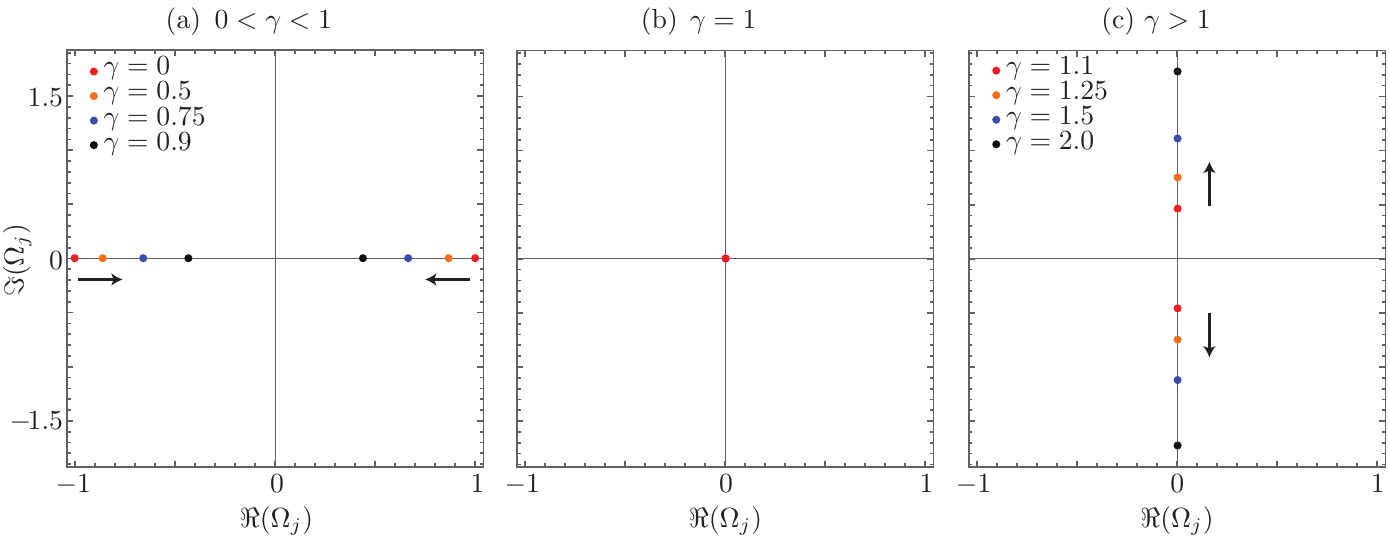}
\caption{Coupling matrix eigenvalue dynamics, (a) $\mathcal{PT}$-symmetric regime, (b) fully degenerate regime, and (c) broken symmetry regime. The black arrows show the direction of the eigenvalues as the gain to coupling ratio increases.}
\label{fig:Figure2}
\end{figure}

Let us try to address propagation in the linear $\mathcal{PT}$-symmetric dimer.
First, note that we can cast its coupled differential equation system, Eq. (\ref{eq:PTDimer}), in a vector form,
\begin{eqnarray}\label{eq:SchrodingerEq}
-i{{\partial }_{\zeta }}\left| \mathcal{E}(\zeta) \right\rangle   =\mathbbm{H}\left| \mathcal{E}(\zeta) \right\rangle,
\end{eqnarray}
similar to Schr\"odinger equation.
Now, the coupling matrix takes the role of a Hamiltonian,
\begin{eqnarray} \label{eq:Hamiltonian}
	   \mathbbm{H}&=& i\gamma {{\sigma }_{z}}+{{\sigma }_{x}},
\end{eqnarray}
where the operators $\sigma_{j}$ with $j=x,y,z$ are Pauli matrices.
In any given case, it is straightforward to find the propagator for this $\zeta$-independent Schr\"odinger-like equation \cite{RodriguezLara2015p5682},
\begin{eqnarray}\label{eq:Propagator}
	\mathbbm{U}\left( \zeta  \right)&=&{{e}^{i\mathbbm{H}\zeta }}, \\
	&=& \cos\left( \Omega \zeta  \right)\mathbbm{1}+\frac{i}{\Omega } \sin\left( \Omega \zeta  \right)\mathbbm{H},
\end{eqnarray}
such that the propagated fields through the device are given in terms of the initial field configuration,
\begin{eqnarray} \label{eq:PropagatedField}
\vert \mathcal{E}(\zeta) \rangle = \mathbb{U}(\zeta) \vert \mathcal{E}(0) \rangle,
\end{eqnarray}
with the dispersion relation given by
\begin{eqnarray} \label{eq:Omega}
	\Omega =\sqrt{1-{{\gamma }^{2}}},
\end{eqnarray}
which can be real for $\gamma < 1$, zero for $\gamma = 1$, or purely imaginary for $\gamma > 1$.
Note that the eigenvalues of the coupling matrix are given by $\pm \Omega$ and they become fully degenerate at $\gamma = 1$.
Figure \ref{fig:Figure2} shows the behavior of the coupling matrix eigenvalues as a function of the gain to coupling ratio as they go from purely real, Fig. \ref{fig:Figure2}(a), degenerate to zero, Fig. \ref{fig:Figure2}(b), and become purely imaginary, Fig. \ref{fig:Figure2}(c).

\begin{figure}[h]
\centering
\includegraphics[width=\textwidth]{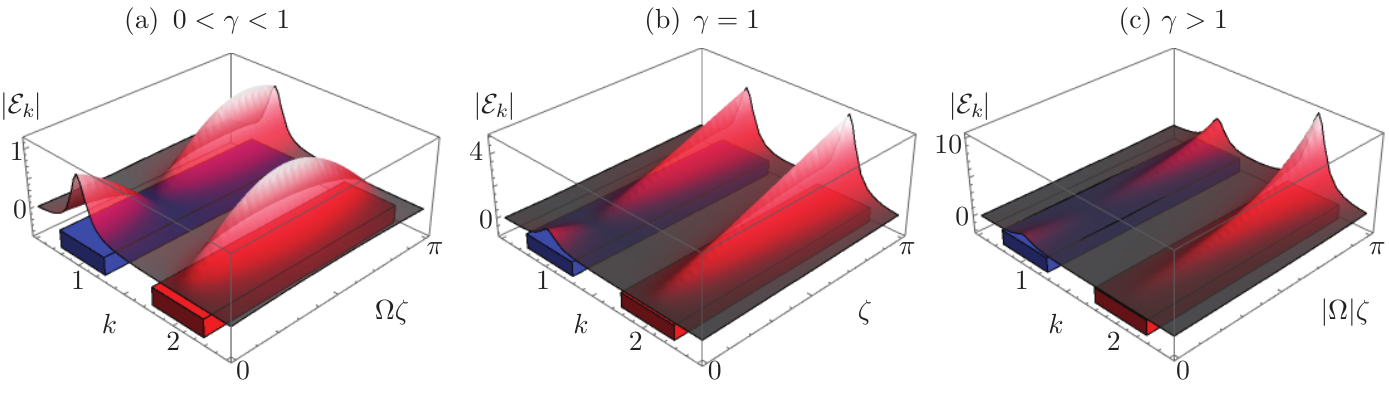}
\caption{Absolute field amplitude propagation in a coupler with effective symmetric loss, blue waveguide, and gain, red waveguide, in the (a) $\mathcal{PT}$-symmetric regime, $\gamma=0.5$, (b) fully degenerate regime, $\gamma=1$, and (c) broken symmetry regime, $\gamma=1.5$.}
\label{fig:Figure3}
\end{figure}

It is straightforward to show that in the $\mathcal{PT}$-symmetric regime, where the eigenvalues of the mode coupling matrix are real, the device behaves as an asymmetric oscillator with amplification,
\begin{eqnarray} \label{eq:PropOsc}
	\mathbbm{U}\left( \zeta  \right)
	&=& \cos\left(  \Omega   \zeta  \right)\mathbbm{1} + \frac{i}{  \Omega  } \sin\left(  \Omega  \zeta  \right)\mathbbm{H}, \quad \gamma < 1.
\end{eqnarray}
Once the $PT$-symmetry is broken, we have two distinct cases, the fully degenerate one where both eigenvalues are zero and the device shows amplification ruled by a power law,
\begin{eqnarray} \label{eq:PropPowAmp}
	\mathbbm{U}\left( \zeta  \right)
	&=& \mathbbm{1} + i  \zeta  \mathbbm{H}, \quad \gamma = 1,
\end{eqnarray}
and the case of purely imaginary eigenvalues, where the amplification is exponential,
\begin{eqnarray} \label{eq:PropExpAmp}
	\mathbbm{U}\left( \zeta  \right)
	&=& \cosh\left( \vert \Omega \vert  \zeta  \right)\mathbbm{1} + \frac{i}{ \vert \Omega \vert } \sinh \left( \vert \Omega \vert \zeta  \right)\mathbbm{H}, \quad \gamma > 1.
\end{eqnarray}
Figure \ref{fig:Figure3} shows the absolute field amplitude propagating through a $\mathcal{PT}$-symmetric dimer when light impinges just at the first waveguide in a device with parameters in the regime with real eigenvalues, Fig. \ref{fig:Figure3}(a), fully degenerate eigenvalues, Fig. \ref{fig:Figure3}(b), and imaginary eigenvalues, Fig. \ref{fig:Figure3}(c).
Now, while we find this algebraic approach short, elegant and elucidating, it is not the only available method to infer the properties of the $\mathcal{PT}$-symmetric dimer.

%%%%%%%%%%%%%%%%%%%%%%%%%%%%%%%%%%%%%%%%%%
\subsection{Nonlocal oscillator, partial differential equation  approach}

It is also possible to decouple the dimer differential equation set, Eq. (\ref{eq:PTDimer}), into two second order differential equations that have the same form of the one-dimensional wave equation \cite{RodriguezLara2015p5682},
\begin{eqnarray}\label{eq:PTDimerOsc}
\partial _{\zeta }^{2}{{\mathcal{E}}_{k}}\left( \zeta  \right)+ \Omega^{2} {{\mathcal{E}}_{k}}\left( \zeta  \right)=0, \quad k=1,2.
\end{eqnarray}
The boundary conditions are given by the initial fields impinging the device, $\mathcal{E}_{k}(0)$, and their first derivatives,
\begin{eqnarray}
	{{\partial }_{\zeta }}{{\mathcal{E} }_{1}}\left( 0  \right)&=& -\gamma {{\mathcal{E}}_{1}}\left( 0 \right)+ i{{\mathcal{E} }_{2}}\left( 0  \right), \\
	{{\partial }_{\zeta }}{{\mathcal{E} }_{2}}\left( 0  \right)&=& \gamma {{\mathcal{E}}_{2}}\left( 0  \right) + i {{\mathcal{E} }_{1}}\left( 0  \right),
\end{eqnarray}
obtained from the dimer mode coupling equation, Eq.(\ref{eq:PTDimer}).
Note that the latter takes us afar from the standard one-dimensional wave equation because it refers to nonlocal media involving linear loss and gain.
Nevertheless, we can follow the standard partial differential equation approach.

In the $\mathcal{PT}$-symmetric case, the characteristic equation is positive, $\Omega^{2}>0$, and we can write the second order differential equation as that of a standard oscillator,
\begin{eqnarray}
\partial _{\zeta }^{2}{{\mathcal{E}}_{k}}\left( \zeta  \right) +  \Omega ^{2} {{\mathcal{E}}_{k}}\left( \zeta  \right)=0,
\end{eqnarray}
thus, we can use its well known solution plus our particular boundary conditions to obtain the propagated fields,
\begin{eqnarray}
{{\mathcal{E}}_{1}}\left( \zeta  \right)&=& {{\mathcal{E}}_{1}}\left(0\right) \cos\left( \vert \Omega  \vert \zeta  \right) - \frac{1}{\vert \Omega \vert } \left[ \gamma {{\mathcal{E}}_{1}}\left(0\right) -i{{\mathcal{E}}_{2}}\left(0\right) \right] \sin\left( \vert \Omega  \vert \zeta  \right)  , \\
{{\mathcal{E} }_{2}}\left( \zeta  \right)&=&{{\mathcal{E}}_{2}}\left(0\right) \cos\left( \vert \Omega \vert \zeta  \right) + \frac{1}{ \vert \Omega  \vert }\left[ \gamma {{\mathcal{E}}_{2}}\left(0\right) + i {{\mathcal{E}}_{1}}\left(0\right) \right] \sin\left( \vert \Omega \vert \zeta  \right).
\end{eqnarray}
These fields allow us to describe the dimer as an asymmetric periodic oscillator with amplification.
It is not an harmonic oscillator due to the boundary condition on the first derivatives.
In the fully degenerate case, where the characteristic equation is equal to zero, $\Omega^{2}=0$,
\begin{eqnarray}\label{eq:PTDimerDFreePart}
\partial _{\zeta }^{2}{{\mathcal{E}}_{k}}\left( \zeta  \right) =0,
\end{eqnarray}
we can think of light propagating through the dimer as a free particle through some nonlocal media with linear unitary gain and loss, that yields amplification following a power law,
\begin{eqnarray}
   {{\mathcal{E}}_{1}}\left( \zeta  \right)&=&{{\mathcal{E}}_{1}}\left(0\right) - \left[ {{\mathcal{E}}_{1}}\left(0\right) - i {{\mathcal{E}}_{2}}\left(0\right)  \right] \zeta , \\
   {{\mathcal{E}}_{2}}\left( \zeta  \right)&=&{{\mathcal{E}}_{2}}\left(0\right) + \left[ {{\mathcal{E}}_{2}}\left(0\right) + i {{\mathcal{E}}_{1}}\left(0\right)  \right] \zeta.
\end{eqnarray}
Finally, in the broken symmetry case, the characteristic equation is negative, $\Omega^{2} < 0$, and we can write the second order differential equation as an inverted oscillator,
\begin{eqnarray}\label{eq:PTDimerOscInv}
\partial _{\zeta }^{2}{{\mathcal{E}}_{k}}\left( \zeta  \right) - \vert \Omega \vert^{2} {{\mathcal{E}}_{k}}\left( \zeta  \right)=0,
\end{eqnarray}
that provides us with a device that amplifies initial fields following an exponential law,
\begin{eqnarray}
{{\mathcal{E}}_{1}}\left( \zeta  \right)&=& {{\mathcal{E}}_{1}}\left(0\right) \cosh\left( \vert \Omega  \vert \zeta  \right) - \frac{1}{\vert \Omega \vert } \left[ \gamma {{\mathcal{E}}_{1}}\left(0\right) -i{{\mathcal{E}}_{2}}\left(0\right) \right] \sinh\left( \vert \Omega  \vert \zeta  \right)  , \\
{{\mathcal{E} }_{2}}\left( \zeta  \right)&=&{{\mathcal{E}}_{2}}\left(0\right) \cosh\left( \vert \Omega \vert \zeta  \right) + \frac{1}{ \vert \Omega  \vert }\left[ \gamma {{\mathcal{E}}_{2}}\left(0\right) + i {{\mathcal{E}}_{1}}\left(0\right) \right] \sinh\left( \vert \Omega \vert \zeta  \right) .
\end{eqnarray}
All these solutions are just the explicit form of the propagated field found earlier by a purely algebraic approach, Eq. (\ref{eq:PropagatedField}) -- Eq. (\ref{eq:PropExpAmp}), and tap into the well known one-dimensional wave equation with the difference that an effective nonlocal active medium is provided by the first derivative boundary conditions.

%%%%%%%%%%%%%%%%%%%%%%%%%%%%%%%%%%%%%%%%%%
\subsection{Nonlinear oscillator, renormalized fields approach}

\begin{figure}[h]
\centering
\includegraphics[width=\textwidth]{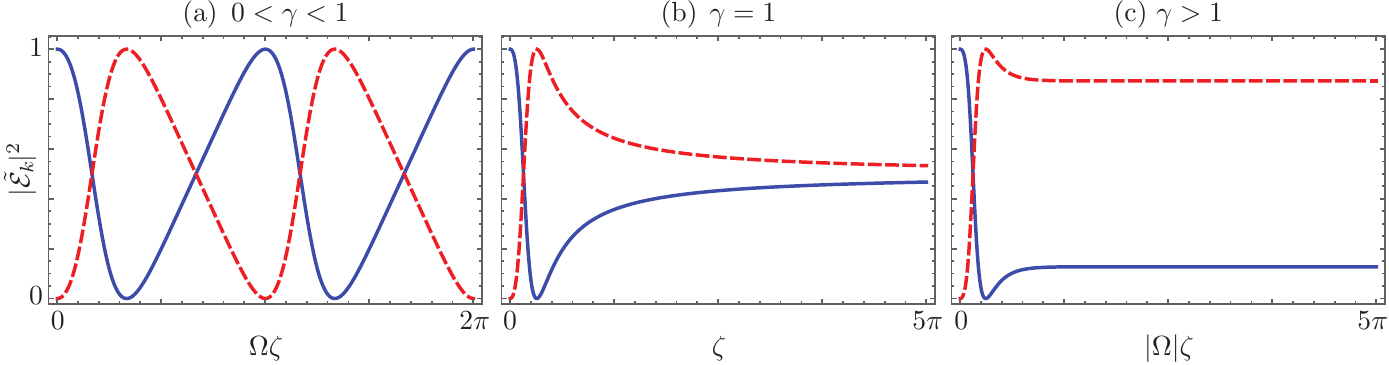}
\caption{Renormalized field intensity propagation in the waveguides with effective loss, $ \vert \tilde{\mathcal{E}}_{1} \vert^{2}$ solid blue line, and gain, $ \vert \tilde{\mathcal{E}}_{2} \vert^{2}$ dashed red line, in the (a) $\mathcal{PT}$-symmetric regime, $\gamma=0.5$, (b) fully degenerate regime, $\gamma=1$, and (c) broken symmetry regime, $\gamma=1.5$, for an initial field impinging just at the first waveguide. }
\label{fig:Figure4}
\end{figure}

So far, we have managed to provide an algebraic propagator and describe the field behavior in the three possible regimes of the linear $\mathcal{PT}$-symmetric dimer.
Now, we can bring forward a complementary view that can give us asymptotic information of the broken symmetry phases.
Let us define instantaneous renormalized fields \cite{RodriguezLara2015p5682},
\begin{eqnarray}
\label{renormalization}
\tilde{\mathcal{E}}_{k}\left( \zeta \right) = \frac{\mathcal{E}_{k}\left( \zeta \right)}{ \sqrt{\vert \mathcal{E}_{1}\left( \zeta \right) \vert^{2} + \vert \mathcal{E}_{2}\left( \zeta \right) \vert^{2}}},
\end{eqnarray}
such that the total renormalized field intensity at each propagation distance is always the unit, $\sum_{k=1}^{2} \vert \tilde{\mathcal{E}}_{k}\left( \zeta \right) \vert^{2} = 1$.
In this picture, it is easier to realize that light intensity through the $\mathcal{PT}$-symmetric device behaves like a nonharmonic oscillator, Fig. \ref{fig:Figure4}(a).
Furthermore, this allows us to conduct asymptotic analysis in the broken symmetry phases.
In the fully degenerate phase, $\gamma = 1$ such that $\Omega = 0$,
it is possible to calculate the asymptotic behavior of the fields intensity as the scaled propagation distance goes to infinity and find out that the renormalized optical power is balanced in both waveguides, independent of the initial field distribution,
\begin{eqnarray}
\label{aymptotic_gamma_equal_1}
\lim_{\zeta \rightarrow \infty} \vert \tilde{\mathcal{E}}_{k}\left( \zeta \right) \vert^{2} = \frac{1}{2}, \quad \gamma = 1, \quad \Omega=0,
\end{eqnarray}
as shown in Fig. \ref{fig:Figure4}(b).
In the broken symmetry regime, where the eigenvalues are purely imaginary, the asymptotic intensity distribution depends on the effective gain to coupling ratio and we can include the previous result,
\begin{eqnarray}
\lim_{\zeta \rightarrow \infty} \vert \tilde{\mathcal{E}}_{1}\left( \zeta \right) \vert^{2} &=& \frac{1}{2\gamma  } \left(\gamma + \sqrt{\gamma^{2}-1}\right)^{-1}, \\
\label{aymptotic_gamma_greater_1_2}
\lim_{\zeta \rightarrow \infty} \vert \tilde{\mathcal{E}}_{2}\left( \zeta \right) \vert^{2} &=& \frac{1}{2\gamma } \left(\gamma + \sqrt{\gamma^{2}-1}\right), \qquad \qquad \gamma \ge 1.
\end{eqnarray}
In other words, for a device long enough, in the broken symmetry region, the input field distribution has no effect on the output field intensity distribution.
The latter is governed only by the gain to coupling ratio of the device, $\gamma$, Fig. \ref{fig:Figure4}(c).

\begin{figure}[h]
\centering
\includegraphics[width=\textwidth]{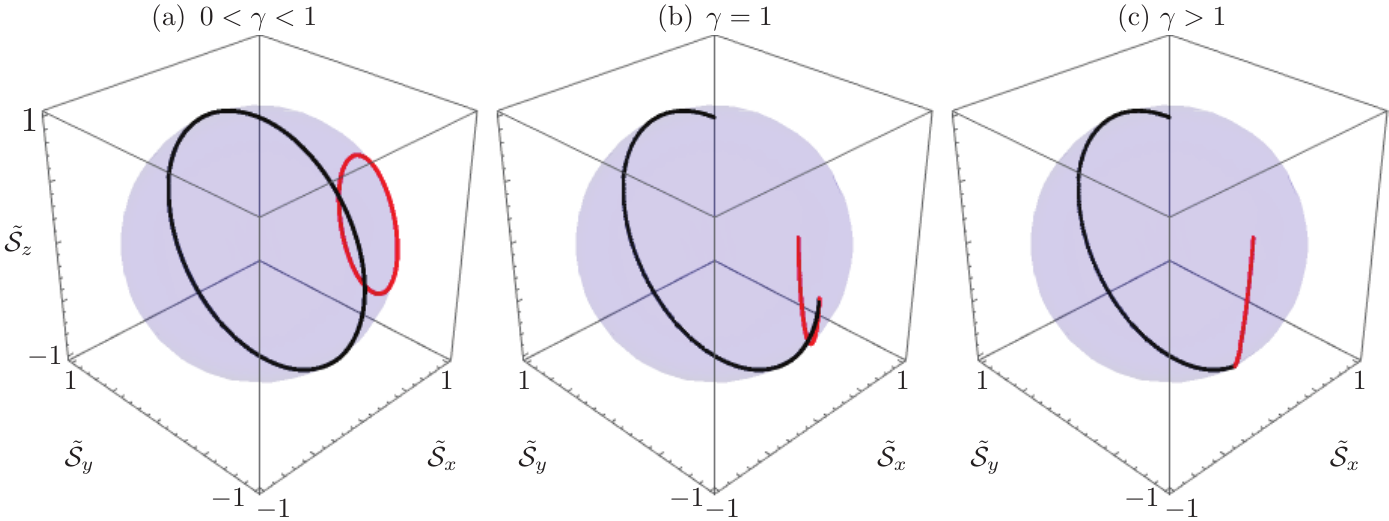}
\caption{Stokes vector propagation in the waveguide coupler with effective symmetric loss and gain, in the (a) $\mathcal{PT}$-symmetric regime, $\gamma=0.5$, (b) fully degenerate regime, $\gamma=1$, and (c) broken symmetry regime, $\gamma=1.5$, for the initial conditions $\mathcal{E}_{1}(0)=1$ and $\mathcal{E}_{2}(0)=0$ in black, and $\mathcal{E}_{1}(0)=1/\sqrt{3}$ and $\mathcal{E}_{2}(0)=\sqrt{1 - \vert \mathcal{E}_{1}(0)\vert^{2}}$ in red. }
\label{fig:Figure5}
\end{figure}

Furthermore, this approach also allows us to show that the linear $\mathcal{PT}$-symmetric dimmer is the irreducible form of two equivalent reducible nonlinear models, one with linear loss and gain, as well as an imaginary nonlinearity,
\begin{eqnarray}
- i \partial_{\zeta} \tilde{\mathcal{E}_{1}}(\zeta)  &=&  \tilde{\mathcal{E}_{2}}(\zeta) + 2 i \gamma \left[ 1- \vert \tilde{\mathcal{E}_{1}}(\zeta) \vert^{2} \right] \tilde{\mathcal{E}_{1}}(\zeta) ,  \\
- i \partial_{\zeta} \tilde{\mathcal{E}_{2}}(z)  &=& \tilde{\mathcal{E}_{1}}(\zeta) - 2 i \gamma \left[ 1- \vert \tilde{\mathcal{E}_{2}}(\zeta) \vert^{2} \right] \tilde{\mathcal{E}_{2}}(\zeta) ,
\end{eqnarray}
and the other corresponding to an imaginary cross-nonlinearity,
\begin{eqnarray}
- i \partial_{\zeta} \tilde{\mathcal{E}_{1}}(\zeta)  &=&  \tilde{\mathcal{E}_{2}}(\zeta) +2 i \gamma \vert \tilde{\mathcal{E}_{2}}(\zeta) \vert^{2} \tilde{\mathcal{E}_{1}}(\zeta) ,  \\
- i \partial_{\zeta} \tilde{\mathcal{E}_{2}}(\zeta)  &=&  \tilde{\mathcal{E}_{1}}(\zeta) -2 i \gamma \vert \tilde{\mathcal{E}_{1}}(\zeta) \vert^{2} \tilde{\mathcal{E}_{2}}(\zeta).
\end{eqnarray}
We can follow a standard approach to deal with nonlinear $\mathcal{PT}$-symmetric dimers \cite{Ramezani2010p043803}, and introduce a Stockes vector, $\vec{\tilde{\mathcal{S}}}=(\tilde{\mathcal{S}}_{x},\tilde{\mathcal{S}}_{y},\tilde{\mathcal{S}}_{z})$, with components given by,
\begin{eqnarray}
\tilde{\mathcal{S}_{j}}(\zeta) = \langle \tilde{\mathcal{E}}(\zeta) \vert \sigma_{j} \vert \tilde{\mathcal{E}}(\zeta) \rangle, \quad j=x,y,z, \label{StockesVector}
\end{eqnarray}
again, the matrices $\sigma_{j}$ are Pauli matrices, the notation $\vert \tilde{\mathcal{E}}(\zeta) \rangle$ is a column vector containing the renormalized field amplitudes equivalent to that defined in Section 2 for the field amplitudes, and the new conjugate transpose notation $\langle \tilde{\mathcal{E}}(\zeta) \vert = \left( \vert \tilde{\mathcal{E}}^{\ast}(\zeta) \rangle\right)^{T}$ is a row vector with the conjugate renormalized field amplitudes as components.
In terms of the renormalized field amplitudes,
\begin{eqnarray}\label{StockesVector2}
 \tilde{\mathcal{S}}_{x}(\zeta) &=& \tilde{\mathcal{E}}^*_{1}(\zeta)\tilde{\mathcal{E}}_{2}(\zeta) +\tilde{\mathcal{E}}_{1}(\zeta) \tilde{\mathcal{E}}^*_{2}(\zeta),  \\
\tilde{\mathcal{S}}_{y}(\zeta)&=& - i \left[\tilde{\mathcal{E}}^*_{1}(\zeta) \tilde{\mathcal{E}}_{2}(\zeta) - \tilde{\mathcal{E}}_{1}(\zeta) \tilde{\mathcal{E}}^*_{2}(\zeta) \right], \\
\tilde{\mathcal{S}}_{z}(\zeta)&=&\vert \tilde{\mathcal{E}}_{1}(\zeta)\vert^2-\vert\tilde{\mathcal{E}}_{2}(\zeta)\vert^2.
\end{eqnarray}
In the case at hand, where the nonlinearity is reducible, the Stokes vector norm is the unit and a constant of motion,
\begin{eqnarray}
 \tilde{\mathcal{S}}(\zeta) = \sqrt{\tilde{\mathcal{S}}^2_{x}(\zeta)+ \tilde{\mathcal{S}}^2_{y}(\zeta) + \tilde{\mathcal{S}}^2_{z}(\zeta)}  = \vert \tilde{\mathcal{E}}_{1}(\zeta) \vert^2 + \vert \tilde{\mathcal{E}}_{2}(\zeta) \vert^2 = 1.
\end{eqnarray}
The Stokes vector components verify the following set of coupled differential equations:
\begin{eqnarray}
\partial_{\zeta}\tilde{\mathcal{S}}(\zeta)&=& 0, \label{eq:RNSV0} \\
\partial_{\zeta}\tilde{\mathcal{S}}_{x}(\zeta)&=& 2\gamma \tilde{\mathcal{S}}_{x}(\zeta)~ \tilde{\mathcal{S}}_{z}(\zeta), \label{eq:RNSVx}\\
\partial_{\zeta}\tilde{\mathcal{S}}_{y}(\zeta)&=&  2 \tilde{\mathcal{S}}_{z}(\zeta) ~ \left[1 +  \gamma \tilde{\mathcal{S}}_{y}(\zeta)\right], \label{eq:RNSVy}\\
\partial_{\zeta}\tilde{\mathcal{S}}_{z}(\zeta)&=& -2 \left\{ \tilde{\mathcal{S}}_{y}(\zeta)  +   \gamma \left[1 -  \tilde{\mathcal{S}}_{z}^2(\zeta) \right] \right\}. \label{eq:RNSVz}
\end{eqnarray}
Note that these equations have a set of stable points only outside the $\mathcal{PT}$-symmetric regime, $\gamma \ge 1$, given by $\tilde{\mathcal{S}}_{s}(\zeta) =(0, -\gamma^{-1}, -\gamma^{-1} \sqrt{\gamma^{2}-1})$.
It is also possible to describe the asymptotic behavior of the renormalized Stokes vector components,
\begin{eqnarray}
\lim_{\zeta \rightarrow \infty} \tilde{\mathcal{S}}_{x}(\zeta)&=& 0, \\
\lim_{\zeta \rightarrow \infty} \tilde{\mathcal{S}}_{y}(\zeta)&=& - \frac{1}{\gamma}, \\
\lim_{\zeta \rightarrow \infty} \tilde{\mathcal{S}}_{z}(\zeta)&=& - \frac{\sqrt{\gamma^2-1}}{\gamma}, \qquad \gamma \ge 1,
\end{eqnarray}
The Stokes vector approach allows us to visualize the field propagation as a trajectory on a unit sphere.
Any initial condition, in the fully degenerate and broken symmetry regimes, will converge asymptotically to the same stable point on the sphere of the differential equation, $\lim_{\zeta \rightarrow \infty} \vec{\tilde{S}}(\zeta) = \tilde{\mathcal{S}}_{s}(\zeta)=(0, -\gamma^{-1}, -\gamma^{-1} \sqrt{\gamma^{2}-1})$, in a dimer described by the gain to coupling ratio $\gamma$.
Figure \ref{fig:Figure5} shows the Stokes vector propagation related to the examples given in Fig. \ref{fig:Figure4} and an additional initial condition set to show that the asymptotic behavior is independent of the initial conditions.
This asymptotical behavior suggest the use of this device as an unidirectional variable amplitude coupler.

%%%%%%%%%%%%%%%%%%%%%%%%%%%%%%%%%%%%%%%%%%
\section{Nonlinear $\mathcal{PT}$-symmetric dimer}

Now, we will discuss a variation of the $\mathcal{PT}$-symmetric dimer where the waveguides have an additional Kerr nonlinearity,
\begin{eqnarray}
- i \partial_{\zeta} \mathcal{E}_{1}(\zeta)  &=&  \left[ i \gamma + \kappa \vert \mathcal{E}_{1}(\zeta) \vert^{2} \right] \mathcal{E}_{1}(\zeta) + \mathcal{E}_{2}(\zeta),  \\
- i \partial_{\zeta} \mathcal{E}_{2}(\zeta)  &=&  \left[ -i \gamma + \kappa \vert \mathcal{E}_{2}(\zeta) \vert^{2} \right] \mathcal{E}_{2}(\zeta) +  \mathcal{E}_{1}(\zeta).
\end{eqnarray}
For the sake of simplicity, we will consider the effective Kerr nonlinearity to coupling strength ratio, $\kappa$, to be equal in both waveguides.
Stability in this system has been formally discussed in the case of $\kappa = 1$ and it was found that the zero equilibrium state is neutrally stable in the $\mathcal{PT}$-symmetric regime, $\gamma < 1$, and that the total light intensity at the waveguides is bounded from above by the initial intensity amplified by an exponential gain, $\vert\mathcal{E}_{1}(\zeta)\vert^{2} + \vert\mathcal{E}_{2}(\zeta)\vert^{2} \le \left( \vert\mathcal{E}_{1}(0)\vert^{2} + \vert\mathcal{E}_{2}(0)\vert^{2} \right)e^{2 \gamma \zeta}$ \cite{Kevrekidis2013p365201}.
In order to deal with the dynamics, it is possible to introduce a Stokes vector defined, now, in terms of the field amplitudes \cite{Ramezani2010p043803},
\begin{eqnarray}
 \mathcal{S}_{x}(\zeta) &=&  {\mathcal{E}}^*_{1}(\zeta) {\mathcal{E}}_{2}(\zeta) + {\mathcal{E}}_{1}(\zeta)  {\mathcal{E}}^*_{2}(\zeta),  \label{eq:SVx}\\
\mathcal{S}_{y}(\zeta)&=& -i \left[ {\mathcal{E}}^*_{1}(\zeta)  {\mathcal{E}}_{2}(\zeta) -  {\mathcal{E}}_{1}(\zeta)  {\mathcal{E}}^*_{2}(\zeta) \right] \label{eq:SVy}, \\
\mathcal{S}_{z}(\zeta)&=& \vert  {\mathcal{E}}_{1}(\zeta)\vert^2-\vert {\mathcal{E}}_{2}(\zeta)\vert^2. \label{eq:SVz}
\end{eqnarray}
Thus, the norm of the redefined Stokes vector, the total intensity at the waveguides, is no longer a constant of motion,
\begin{eqnarray}
 \mathcal{S}(\zeta) = \sqrt{\mathcal{S}_{x}^2(\zeta)+ \mathcal{S}_{y}^2(\zeta) + \mathcal{S}_{z}^2(\zeta)} = \vert  {\mathcal{E}}_{1}(\zeta) \vert^2 + \vert  {\mathcal{E}}_{2}(\zeta) \vert^2, \label{eq:SV0}
\end{eqnarray}
and its dynamics,
\begin{eqnarray}
\partial_{\zeta} \mathcal{S}\left( \zeta \right) &=& - 2 \gamma \mathcal{S}_{z}\left( \zeta \right), \label{eq:SVNL0}\\
\partial_{\zeta} \mathcal{S}_{x}\left( \zeta \right) &=&  \kappa \mathcal{S}_{y}\left( \zeta \right) \mathcal{S}_{z} \left( \zeta \right), \label{eq:SVNLx}\\
\partial_{\zeta} \mathcal{S}_{y}\left( \zeta \right) &=&   \mathcal{S}_{z}\left( \zeta \right) \left[ 2 - \kappa \mathcal{S}_{x} \left( \zeta \right) \right] , \label{eq:SVNLy} \\
\partial_{\zeta} \mathcal{S}_{z}\left( \zeta \right) &=&  - 2 \left[\mathcal{S}_{y} \left( \zeta \right) + \gamma \mathcal{S}\left( \zeta \right)   \right] , \label{eq:SVNLz}
\end{eqnarray}
will not be restricted to the unit sphere.
These dynamics have been shown to be an optical simulation of a relativistic massless particle of negative charge in a pseudolectromagnetic field \cite{Ramezani2010p043803}.
Note that there is a set of stable points, $\partial_{\zeta} \vec{S}_{s} (\zeta) = 0$, for the effective gain to coupling ratio $\gamma < 1$,
\begin{eqnarray}
\vec{S}_{s}(\zeta) &=& \left\{ \left( \frac{2}{\kappa}~,~ \pm \frac{ 2 \gamma}{ \kappa \sqrt{1 -\gamma^{2}}}~,~ 0 \right) , \right. \nonumber \\
&& \left. \left( \pm \frac{1}{\gamma} \sqrt{(1 - \gamma^2)}~ \vert {S}_{y}(\zeta) \vert~,~~ \mathcal{S}_{y}(\zeta)~,~ 0 \right)  \right\} .
\end{eqnarray}
All these stable nonlinear modes are waveguide fields of the form $\mathcal{E}_{j} = \mathcal{A} e^{i \phi _{j}}$ with $\mathcal{A}\ge 0$ and phase difference constricted by the relation,
\begin{eqnarray}
\tan \left(\phi_{1} - \phi_{2}\right)  = \pm \frac{\gamma}{\sqrt{1-\gamma^{2}}}.
\end{eqnarray}
There are no stable points outside the $\mathcal{PT}$-symmetric regime, $\gamma \ge 1$.

\begin{figure}[h]
\centering
\includegraphics[width=\textwidth]{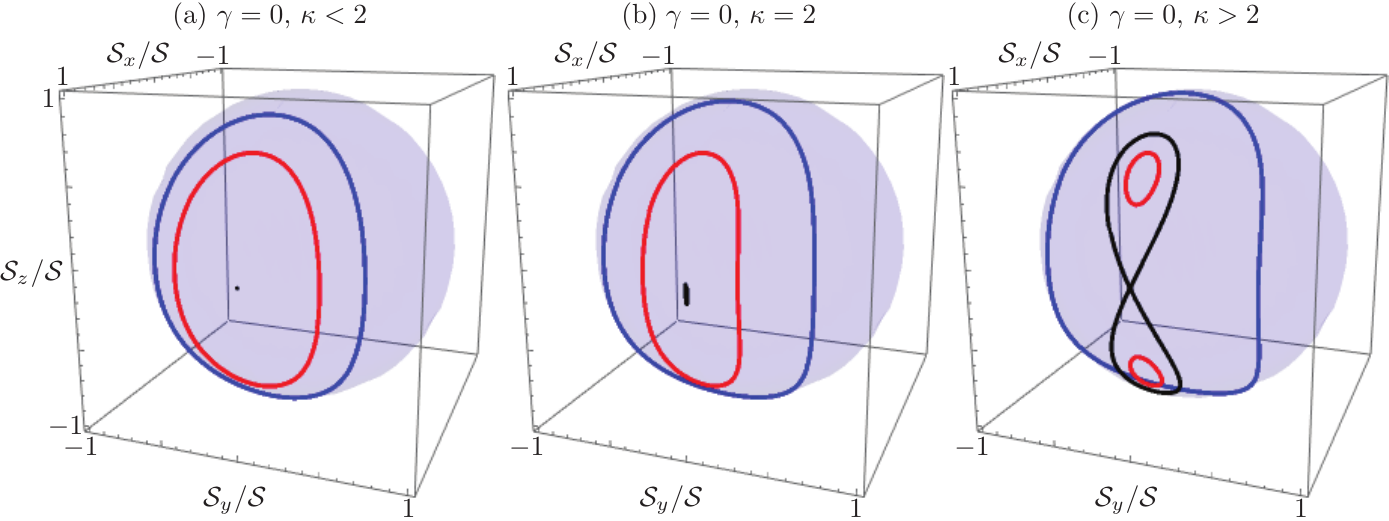}
\caption{Stokes vector propagation in the passive two-waveguide coupler, $\gamma=0$, with nonlinearities (a) below, $\kappa=1.5$, (b) at, $\kappa=2$, and (c) above, $\kappa=2.5$, the critical Kerr nonlinearity to coupling strength ratio, $\kappa=2$. The figure shows (a) stationary point, (b) trajectory infinitesimally near the stationary point, and (c) separatrix  in black, (a)-(b) Rabi and (c) Josephson oscillations in red, and (a)-(c) Rabi oscillations in blue. }
\label{fig:Figure6}
\end{figure}

In order to create intuition, let us start with the passive self-focusing two-waveguide coupler, $\gamma = 0$, that has two constants of motion in the form of the Stokes vector norm, $S(\zeta)=1$ such that $\partial_{\zeta} S(\zeta) = 0$, and the Hamiltonian-like quantity,
\begin{eqnarray}
\mathcal{H} = \frac{1}{2} \kappa \mathcal{S}_{z}^2(\zeta) + 2 \mathcal{S}_{x}(\zeta).
\end{eqnarray}
Note that the system is integrable and the conservation of the Stokes vector norm allows for the parametrization $\mathcal{S}_{x}(\zeta)= \sqrt{1 - \mathcal{S}_{z}^{2}} ~\cos \phi (\zeta)$ and $\mathcal{S}_{y}(\zeta)= \sqrt{1 - \mathcal{S}_{z}^{2}} ~\sin \phi (\zeta)$, such that we can write a Hamilton-Jacobi model,
\begin{eqnarray}
&&\mathcal{H} = \frac{1}{2}\kappa \mathcal{S}_{z}^2(\zeta) + 2 \sqrt{1 - \mathcal{S}_{z}^{2}(\zeta)} \cos \phi (\zeta), \qquad \nonumber \\
&&\partial_{\zeta} \mathcal{S}_{z}(\zeta) = \frac{\partial \mathcal{H}}{\partial \phi} , \qquad
\partial_{\zeta} \phi (\zeta) = - \frac{\partial \mathcal{H}}{\partial \mathcal{S}_{z}},
\end{eqnarray}
equivalent to that of a nonrigid pendulum or a Bose-Josephson junction \cite{Smerzi1997p4950,RodriguezLara2011p016225}.
This particular configuration allows for Rabi oscillations, Fig. \ref{fig:Figure6}(a), below the critical effective Kerr nonlinearity to coupling ratio, $\kappa = 2$, Fig. \ref{fig:Figure6}(b), above this critical value the system can show both Rabi and Josephson oscillations, Fig. \ref{fig:Figure6}(c).
In other words, the initial field amplitudes either coherently oscillate between the waveguides or localize at the waveguide where they were originally prepared, depending on both the initial field distribution and the effective Kerr nonlinearity of the device.

\begin{figure}[h]
\centering
\includegraphics[width=\textwidth]{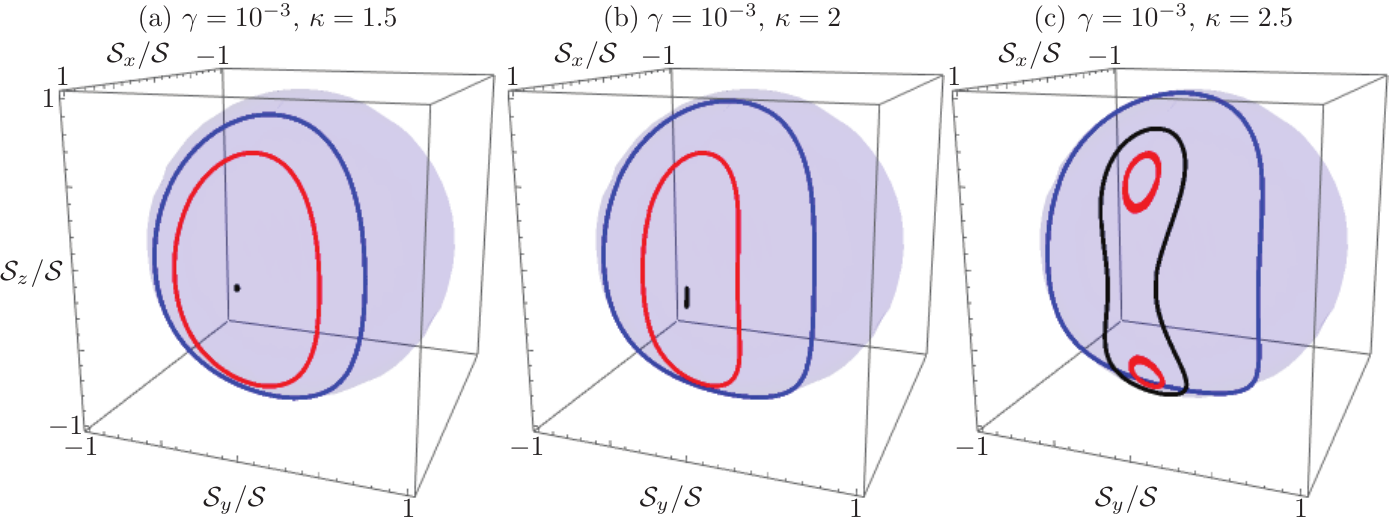}
\caption{Renormalized Stokes vector propagation in the waveguide coupler with a fixed gain to coupling ratio $\gamma=0.001$  and variable effective nonlinearity to coupling ratio (a) $\kappa=1.5$, (b) $\kappa=2$, and (c) $\kappa=2.5$, for the same initial conditions than Fig. \ref{fig:Figure6}. }
\label{fig:Figure7}
\end{figure}

If we now include the symmetric gain and loss, $\gamma \neq 0$, the system is still integrable as we can write two constants of motion \cite{Ramezani2010p043803},
\begin{eqnarray}
\mathcal{S}_{C}\left( \zeta \right) &=& \sqrt{ \left[\kappa \mathcal{S}_{x}\left( \zeta \right) -2 \right]^{2} + \left[ \kappa \mathcal{S}_{y}\left( \zeta \right) \right]^{2}}, \\
\mathcal{S}_{J}\left( \zeta \right) &=& S\left( \zeta \right) + \frac{2 \gamma}{\kappa} \arctan \left[ \frac{ \kappa \mathcal{S}_{x} \left( \zeta \right) - 2}{  \kappa \mathcal{S}_{y}\left( \zeta\right)} \right].
\end{eqnarray}
In the literature, it has been found that the system is stable in the interval $0 \le \gamma <1$ for effective nonlinearity $\kappa = 1$ \cite{Kevrekidis2013p365201} and numerical arguments have been given in the most general case \cite{Ramezani2010p043803}.
Figure \ref{fig:Figure7} shows how the dynamics of the passive nonlinear dimer are affected by the addition of a small effective gain to coupling ratio to the system.
Below the critical nonlinearity for the passive system, Fig. \ref{fig:Figure7}(a), we can still find the coherent oscillation behavior of the linear $\mathcal{PT}$-symmetric dimer but the former stable point is no longer a fixed point of the system.
As the nonlinearity increases, we can see how device parameters and initial conditions start having an effect on the dynamics, Fig. \ref{fig:Figure7}(b), until a point where it is possible to have unstable light localization at the waveguide where light originally impinged, Fig. \ref{fig:Figure7}(c).
Here, the constant of motion plays an important role as an accuracy test for the process of numerically solving the coupled nonlinear system.

\begin{figure}[h]
\centering
\includegraphics[width=\textwidth]{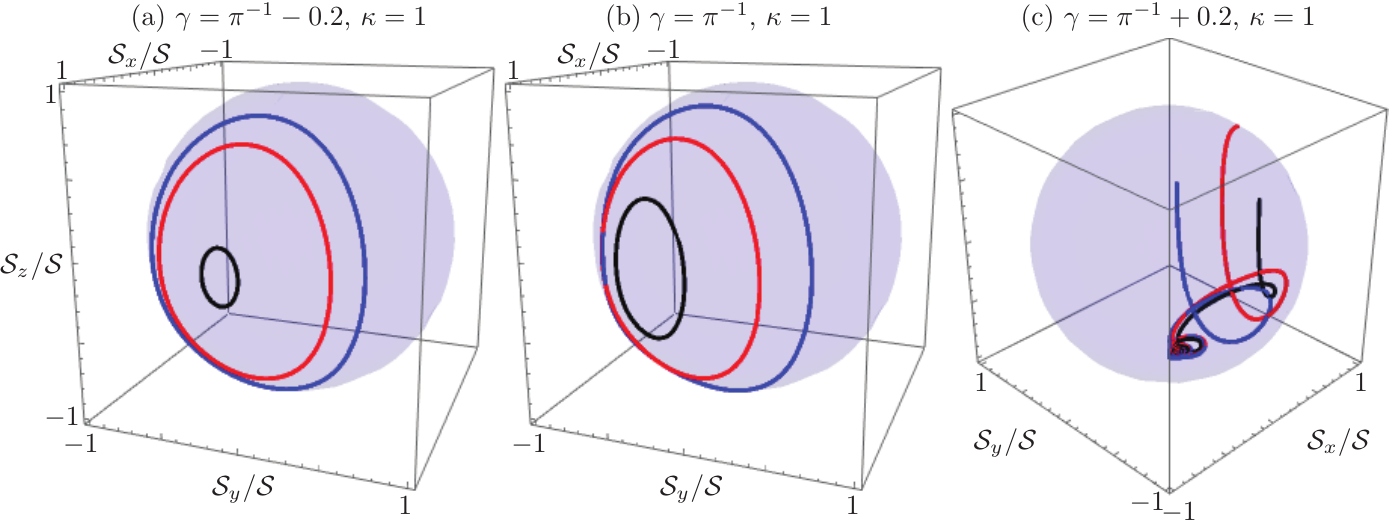}
\caption{Renormalized Stokes vector propagation in the waveguides waveguide coupler with a fixed effective Kerr nonlinearity to coupling ratio $\kappa=1$  and variable effective gain to coupling ratio (a) $\gamma= \pi{-1} -0.2$, (b) $\gamma=\pi^{-1}$, and (c) $\gamma=\pi^{-1}+0.2$, for the same initial conditions than Fig. \ref{fig:Figure7}. }
\label{fig:Figure8}
\end{figure}

Now, for a given effective Kerr nonlinearity to coupling ratio, $\kappa = 1$, we can approximate a critical effective gain to coupling ratio, $\gamma_{c} \simeq \pi^{-1} $, where a change of dynamics occur.
First, the reciprocity condition regarding the exchange of waveguides for a given initial field distribution breaks as we get closer to the approximate critical gain to coupling ratio, note the shift to the left in Fig. \ref{fig:Figure8}(a) and \ref{fig:Figure8}(b), then, after we cross the critical value, the field intensity at the gain waveguide gets localized and experiences an exponential gain, while the field in the lossy waveguide diminishes independently of the initial field amplitude distribution, Fig. \ref{fig:Figure8}(c).
Note, the dynamics above the critical gain to coupling ratio do not tend to the fixed point of the system as in the linear device; the fields do not seem to show a constant behavior in the asymptotic limit.
This localization with amplification in the gain waveguide suggests the use of these devices as optical diodes \cite{Ramezani2010p043803}.

%%%%%%%%%%%%%%%%%%%%%%%%%%%%%%%%%%%%%%%%%%
\section{Linear $\mathcal{PT}$-symmetric planar $N$-waveguide coupler}

\begin{figure}[h]
\centering
\includegraphics[width=\textwidth]{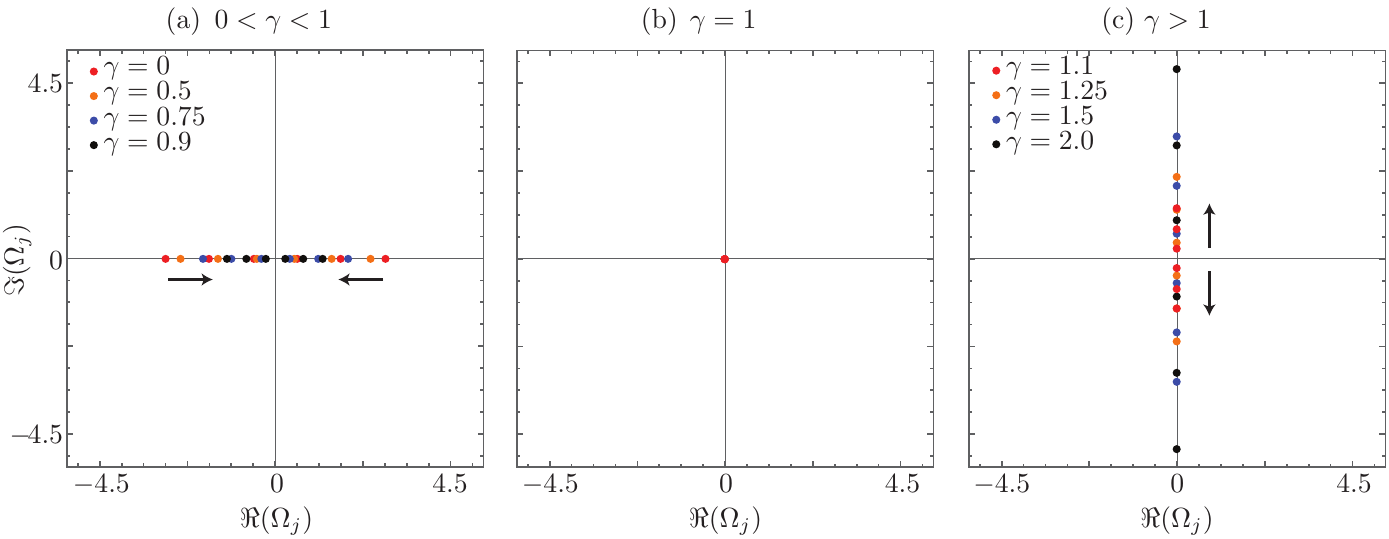}
\caption{Coupling matrix eigenvalue dynamics, (a) $\mathcal{PT}$-symmetric regime, (b) fully degenerate regime, and (c) broken symmetry regime. The black arrows show the direction of the eigenvalues as the gain to coupling ratio increases. Theses cases show the results for a $N=6$ waveguide coupler that provides a Bargmann parameter $j=5/2$.}
\label{fig:Figure9}
\end{figure}

As we said before, the linear dimer is the device with the smallest dimension showing $\mathcal{PT}$-symmetry.
Its algebraic structure, Eq.(\ref{eq:Hamiltonian}), and the fact that Pauli matrices are the two by two scaled matrix representation of the $SU(2)$ group in dimension two, $\sigma_{k} = 2 J_{k}$, suggest that a larger class of $N$-waveguide devices can be constructed with higher dimensional matrix representations of $SU(2)$.
Therefore, we can construct a mode coupling equation set describing a device of $N$ waveguides following Eq. (\ref{eq:PTDimer}),
\begin{eqnarray}
i \partial_{z} \vert \mathcal{E}(z) \rangle = 2 \left( i n_{-} \mathbbm{J}_{z} + g \mathbbm{J}_{x} \right) \vert \mathcal{E}(z) \rangle.
\end{eqnarray}
Now, we choose to introduce a different scaled propagation, $\xi = 2 \zeta = 2gz$, to recover a single parameter Schr\"odinger-like equation with the form,
\begin{eqnarray} \label{eq:HSU2}
	i \partial_{\xi} \vert \mathcal{E}(\xi) \rangle = \mathbbm{H} \vert \mathcal{E}(\xi) \rangle, \qquad \mathbbm{H}= i \gamma \mathbbm{J}_{z}+ \mathbbm{J}_{x} ,
\end{eqnarray}
where the effective refractive index to coupling ratio is the same that in the standard dimer, $\gamma =  n_{-} / g$.
This mode coupling matrix with $i \gamma \in \mathbb{R}$, describes planar $N$-waveguide couplers with identical real effective refractive indices and underlying $SU(2)$ symmetry that show harmonic oscillator behavior \cite{VillanuevaVergara2015p}.
These devices have been used to produce perfect state transfer in both the classical and quantum regimes of optical circuits \cite{PerezLeija2013p012309,ElGanainy2013p161105,RodriguezLara2014p013802,Chapman2016p11339}.
In order to deal with a linear $\mathcal{PT}$-symmetric $N$-waveguide coupler, we must consider pure imaginary effective refractive indices, $\gamma \in \mathbb{R}$, and, thus, the realization of higher finite-dimensional non-unitary representation of $SO(2,1)$, implemented as a complexified version of $SU(2)$, are involved.
In matrix form, the group generators have the following elements,
\begin{eqnarray}
\left[ \mathbbm{J}_{x}\right]_{m,n} &=& \frac{1}{2} \left[\delta_{m-1,n} \sqrt{ n \left(2j-n+1\right)} ~  + \delta_{m+1,n} \sqrt{ m \left(2j-m+1\right)}   \right],\\
\left[ \mathbbm{J}_{y}\right]_{m,n} &=& \frac{i}{2} \left[\delta_{m-1,n} \sqrt{ n \left(2j-n+1\right)} ~ - \delta_{m+1,n} \sqrt{ m \left(2j-m+1\right)}   \right], \\
\left[ \mathbbm{J}_{z}\right]_{m,n} &=& \delta_{m,n} \left( j-m+1 \right), \qquad \qquad \qquad m,n = 1,2,\ldots, N,
\end{eqnarray}
with the Kronecker delta given by $\delta_{m,n}$ and the Bargmann parameter by $j=(N-1)/2$.
These matrices fulfill the commutation relation $\left[ \mathbbm{J}_{i}, \mathbbm{J}_{j} \right] = i \epsilon_{ijk} \mathbbm{J}_{z}$, where $\epsilon_{ijk}$ is the Levi-Civitta symbol, and commute with the Casimir operator $\mathbbm{J}^{2} = \mathbbm{J}_{x}^{2}+\mathbbm{J}_{y}^{2}+\mathbbm{J}_{z}^{2} = j(j+1) \mathbbm{1}$, $\left[\mathbbm{J}_{j}, \mathbbm{J}^{2}\right]=0$.
In the standard differential form, this is equivalent to the coupled mode set,
\begin{eqnarray}
 - i \partial_{\xi} \mathcal{E}_{k}(\xi) &=& \frac{i \gamma}{2} \left(N-2k+1 \right) ~ \mathcal{E}_{k}(\xi) + \nonumber \\
  && \frac{1}{2} \sqrt{ (k-1)(N-k-1)} \mathcal{E}_{k-1}(\xi) + \nonumber \\
   && \frac{1}{2} \sqrt{ k (N-k)} ~ \mathcal{E}_{k+1}(\xi).
\end{eqnarray}
Following the vector notation, we can construct a field vector as,
\begin{eqnarray}
\vert \mathcal{E}(\xi) \rangle = \sum_{k=1}^{N} \mathcal{E}_{k}(\xi) \vert j, j-k+1 \rangle,
\end{eqnarray}
where we can define the $n$th element of the standard basis as
\begin{eqnarray}
\left[ \vert j, m \rangle \right]_{n} = \delta_{j-m+1,n}\,, \qquad m= -j, -j+1, \ldots, j-1, j,
\end{eqnarray}
such that we can define more helpful generators with their corresponding actions,
\begin{eqnarray}
J_{z}  \vert j,m \rangle &=&  ~m ~\vert j,m \rangle, \\
J_{\pm}  \vert j, m \rangle &=& \sqrt{(m+1)(2j + 1 \mp m)} ~  ~ \vert j,m \pm 1 \rangle,
\end{eqnarray}
where we have defined the ladder operators $\mathbbm{J}_{\pm} = \mathbbm{J}_{x} \pm i  \mathbbm{J}_{y}$ that fulfill $\left[ \mathbbm{J}_{z}, \mathbbm{J}_{\pm} \right] = \pm \mathbbm{J}_{\pm}$.
Following the Gilmore-Perelomov approach for $SU(2)$  \cite{VillanuevaVergara2015p,RodriguezLara2015p5682}, we can find the $n$th eigenvalue of the mode coupling matrix,
\begin{eqnarray}
\Omega_{n} = (j-n+1) ~ \Omega, \qquad n=1,\ldots,N
\end{eqnarray}
and obtain the same structure found for the dimer.
All the eigenvalues will be real numbers for $\gamma <1$, Fig. \ref{fig:Figure9}(a), completely degenerate and equal to zero $\Omega_{m} = 0$ for $\gamma =1$, Fig. \ref{fig:Figure9}(b), and imaginary for $\gamma >1$, Fig. \ref{fig:Figure9}(c).
This so-called collapse of the eigenvalues is a direct consequence of the underlying symmetry.

\begin{figure}[h]
\centering \includegraphics[width=\textwidth]{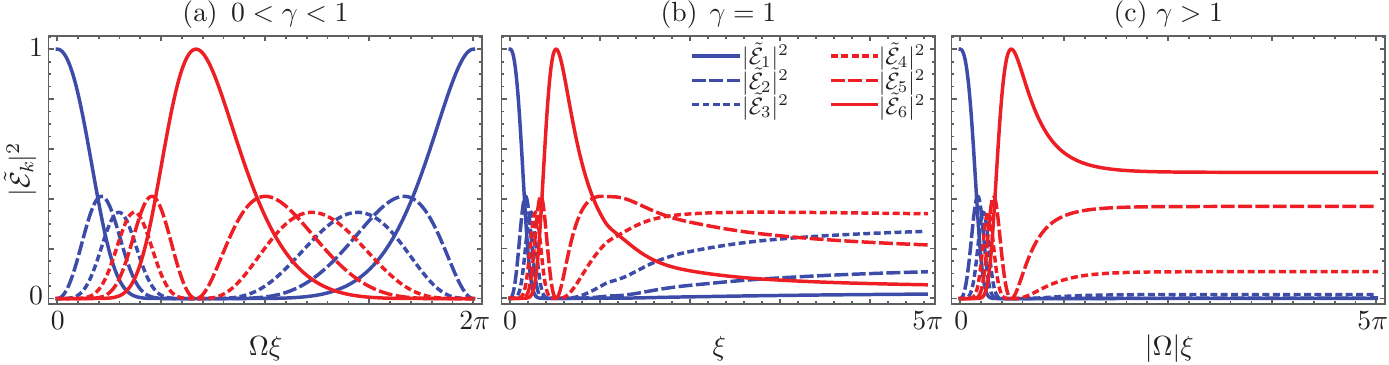}
\caption{Renormalized field intensity propagation for a $N=6$ waveguide coupler, Bargmann parameter $j=5/2$, in the (a) $\mathcal{PT}$-symmetric regime, $\gamma=0.5$, (b) fully degenerate regime, $\gamma=1$, and (c) broken symmetry regime, $\gamma=1.5$, for an initial field impinging just the first waveguide.}
\label{fig:Figure10}
\end{figure}

Furthermore, we can provide the propagation matrix elements,
\begin{eqnarray}
\left[ \mathbb{U}(\xi) \right]_{m,n}&=&  \Omega^{-2j} \sqrt{ \left( \begin{array}{c} 2j \\ m-1 \end{array} \right) \left( \begin{array}{c} 2j \\ n-1 \end{array}\right) } ~   \nonumber \\
&& \left(  \Omega \cos \frac{\Omega}{2} \xi -  \gamma \sin \frac{\Omega}{2} \xi \right)^{2(j+1)-m-n}  \left(  i  \sin \frac{\Omega}{2} \xi  \right)^{m+n-2}  \nonumber \\
&& ~_{2}F_{1} \left( 1-m,1-n,-2j, \Omega^{2}  \csc^{2}  \frac{\Omega}{2} \xi  \right), \qquad \gamma <1,  \label{eq:ProgSU2a}\\
\left[ \mathbb{U}(\xi) \right]_{m,n}&=& (-2)^{-2j} ~  \sqrt{ \left( \begin{array}{c} 2j \\ m-1 \end{array} \right) \left( \begin{array}{c} 2j \\ n-1 \end{array}\right) } ~  \left( \xi - 2 \right)^{2(j+1)-m-n} ~ \nonumber \\
&& \left( i \xi \right) ^{m+n-2}  ~_{2}F_{1} \left( 1-m,1-n,-2j,   \frac{4}{\xi^{2}}    \right), \qquad \gamma =1, \label{eq:ProgSU2b}\\
\left[ \mathbb{U}(\xi) \right]_{m,n}&=& \left(\vert \Omega\vert  \right)^{-2j} \sqrt{ \left( \begin{array}{c} 2j \\ m-1 \end{array} \right) \left( \begin{array}{c} 2j \\ n-1 \end{array}\right) } ~   \nonumber \\
&& \left( \vert\Omega\vert \cosh \frac{\vert\Omega\vert}{2} \xi -  \gamma \sinh \frac{\vert\Omega\vert}{2} \xi \right)^{2(j+1)-m-n}  \left(  i \sinh \frac{\vert\Omega\vert}{2} \xi  \right)^{m+n-2}  \nonumber \\
&& ~_{2}F_{1} \left( 1-m,1-n,-2j, \vert \Omega \vert ^{2}  \mathrm{csch}^{2}  \frac{\vert\Omega\vert}{2} \xi  \right), \qquad \gamma >1, \label{eq:ProgSU2c}
\end{eqnarray}
where the notation $\left( \begin{array}{c} a \\ b \end{array} \right)$ and $_{2}F_{1}(a,b,c,z)$ stand for binomial coeffcient and Gauss hypergeometric function, in that order.
Again, we will have three distinct propagation behaviors as demonstrated for the dimer.
These behaviors are simpler to visualize if we define renormalized field amplitudes,
\begin{eqnarray}
\tilde{\mathcal{E}}_{k}\left( \xi \right) = \frac{\mathcal{E}_{k}\left( \xi \right)}{ \sqrt{ \sum_{p=1}^{N} \vert \mathcal{E}_{p}\left( \xi \right) \vert^{2} }}.
\end{eqnarray}
Now, we can see periodical amplified oscillations in the $\mathcal{PT}$-symmetric regime, $\gamma < 1$, Fig. \ref{fig:Figure10}(a) , amplification following a power law in the fully degenerate regime, $\gamma = 1$, Fig. \ref{fig:Figure10}(b), asymmetric amplification following an exponential law in the broken symmetry regime, $\gamma > 1$, Fig. \ref{fig:Figure10}(c) .

Also, we can derive the set of reducible coupled nonlinear equations for the renormalized field amplitudes,
\begin{eqnarray}
-i \partial \tilde{\mathcal{E}}_{k}\left( \xi \right) &=& \frac{i\gamma}{2} \left[  \left( N-2k+1 \right) +\sum_{p=1}^{N} \left( N -2p + 1\right) \vert \tilde{\mathcal{E}}_{p}\left( \xi \right) \vert^{2} \right] \tilde{\mathcal{E}}_{k}\left( \xi \right) + \nonumber \\
&& + \frac{1}{2} \sqrt{ (k-1)(N-k-1)}~ \tilde{\mathcal{E}}_{k-1}\left(\xi\right) +  \nonumber \\
&&  \frac{1}{2} \sqrt{ k (N-k)} ~ \tilde{\mathcal{E}}_{k+1}\left(\xi\right), \quad k=1,2, \ldots, N.
\end{eqnarray}
It is cumbersome but possible to show that the asymptotic response of a $N$-waveguide coupler in the fully degenerate and broken symmetry regimes is independent of the input field amplitude distribution,
\begin{eqnarray}
\label{eq:AsympNPlanar}
\lim_{\xi \rightarrow \infty} \vert \tilde{\mathcal{E}}_{k}\left( \xi \right) \vert^{2} &=&  \frac{1}{\left(2 \gamma \right)^{2j}}   \left( \begin{array}{c} 2j \\ k-1 \end{array} \right)  \left( \gamma + \sqrt{\gamma^2 - 1} \right)^{2\left( k-j-1 \right)}, ~ \gamma \ge 1.
\end{eqnarray}
which can be seen in Fig. \ref{fig:Figure10}(b) for $\gamma = 1$ and Fig. \ref{fig:Figure10}(c) for $\gamma >1$.
As expected from the mathematical description, in the fully degenerate regime, Fig. \ref{fig:Figure10}(b), the extremal waveguides, those with major effective losses and gain, will transmit fields with smaller amplitudes than those in the central waveguides, because the asymptotic intensity distribution follows the binomial coefficient.
In the broken symmetry regime, Fig. \ref{fig:Figure10}(c), the field intensity correlates with the strength of the gain or loss; the most intense field will travel through the waveguide with the larger effective gain, and the less intense through the one with the larger effective loss.

\begin{figure}[h]
\centering
\includegraphics[width=\textwidth]{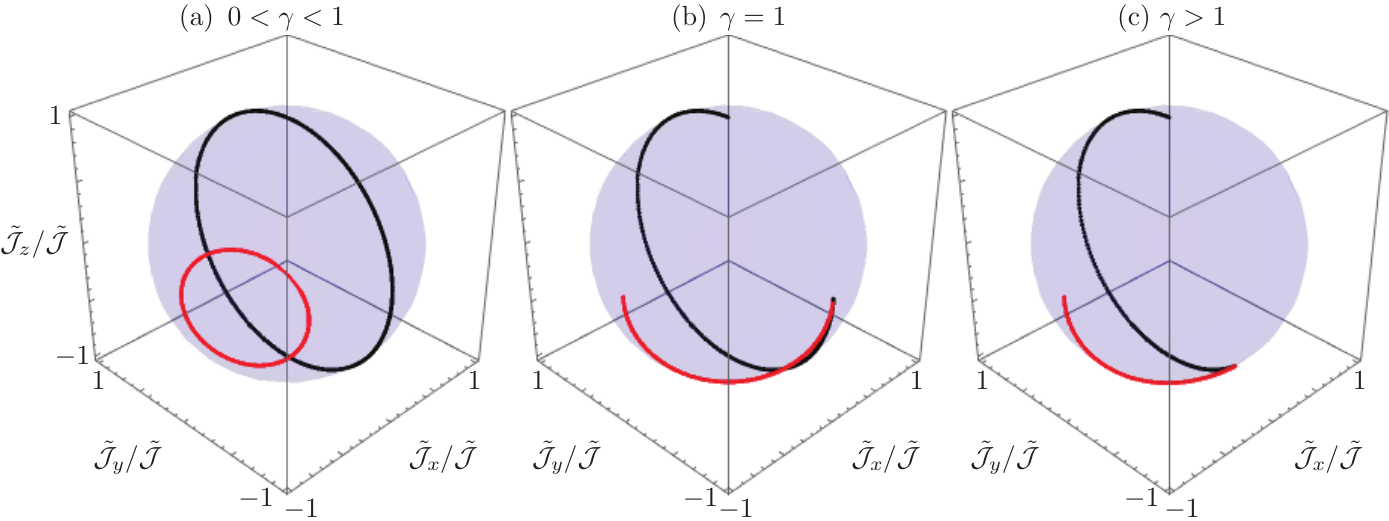}
\caption{Renormalized Stokes vector propagation in a six-waveguide coupler, $j=5/2$, in the (a) $\mathcal{PT}$-symmetric regime, $\gamma=0.5$, (b) fully degenerate regime, $\gamma=1$, and (c) broken symmetry regime, $\gamma=1.5$, for the initial conditions $\mathcal{E}_{k}(0)=\delta_{k,1}$ in black, and the  eigenstate of $J_{x}$ with eigenvalue $-j$, $J_{x} \vert \mathcal{E}(0) \rangle = -j \vert \mathcal{E}(0) \rangle $, in red.}
\label{fig:Figure11}
\end{figure}

The equivalent three-dimensional formulation of the Stokes vector for renormalized fields,
\begin{eqnarray}
\tilde{\mathcal{J}}_{k}( \xi ) = \langle \tilde{\mathcal{E}}\left( \xi \right) \vert \mathbbm{J}_{k} \vert \tilde{\mathcal{E}}\left( \xi \right) \rangle, \qquad k=x,y,z, \label{NStockesVector}
\end{eqnarray}
yields components of the following form,
\begin{eqnarray}
\tilde{\mathcal{J}}_{x}\left(\xi\right) &=& \frac{1}{2} \sum_{k=1}^{2j+1} \sqrt{ k(2j-k+1)} \left[ \tilde{\mathcal{E}}_{k+1}^{\ast} \tilde{\mathcal{E}}_{k} + \tilde{\mathcal{E}}_{k}^{\ast} \tilde{\mathcal{E}}_{k+1} \right], \\
\tilde{\mathcal{J}}_{y}\left(\xi\right) &=& \frac{i}{2} \sum_{k=1}^{2j+1} \sqrt{ k(2j-k+1)} \left[ \tilde{\mathcal{E}}_{k+1}^{\ast} \tilde{\mathcal{E}}_{k} - \tilde{\mathcal{E}}_{k}^{\ast} \tilde{\mathcal{E}}_{k+1} \right], ,\\
\tilde{\mathcal{J}}_{z}\left(\xi\right) &=& \sum_{k=1}^{2j+1} \left(j-k+1 \right) \vert \tilde{\mathcal{E}}_{k}\left( \xi \right) \vert^{2}.
\end{eqnarray}
Here, the conserved variable is the Casimir operator,
\begin{eqnarray}
\tilde{\mathcal{C}}(\xi) &=& \langle \tilde{\mathcal{E}}\left( \xi \right)  \vert \left[ \mathbbm{J}_{x}^{2} + \mathbbm{J}_{y}^{2} + \mathbbm{J}_{z}^{2} \right] \vert \tilde{\mathcal{E}}\left( \xi \right) \rangle = j(j+1),
\end{eqnarray}
and it is important to emphasize that the norm of this three-dimensional Stokes vector is no longer a constant of motion,
\begin{eqnarray}
\tilde{\mathcal{J}}(\xi) = \sqrt{\tilde{\mathcal{J}}_{x}^{2}(\xi)+\tilde{\mathcal{J}}_{y}^{2}(\xi)+\tilde{\mathcal{J}}_{z}^{2}(\xi)}.
\end{eqnarray}
The reason behind this is that a complex vector of dimension $N$ with unit norm can be represented as a point on  the surface of a ball of unit radius in dimension $N^{2}-1$.
For example, the renormalized fields through a two-waveguide coupler, two-dimensional complex vector of unit norm, can be represented on the surface of a three-dimensional ball; in other words, a two-dimensional sphere, where
the Stokes vector norm is a constant of motion,
\begin{eqnarray}
 \tilde{\mathcal{J}}(\xi) = j \left[ \vert \tilde{\mathcal{E}}_{1}\left( \xi \right) \vert^{2} + \vert \tilde{\mathcal{E}}_{2}\left( \xi \right) \vert^{2} \right] = \frac{1}{2}, \qquad j=\frac{1}{2}.
\end{eqnarray}
In general, we should use the surface of a $(N^2-1)$-ball of unit radius in order to describe properly the field amplitudes propagating through a $N$-waveguide coupler.
This does not make it simpler to visualize the dynamics, so, we favor a projection from $(N^{2}-1)$-dimensional to three-dimensional space with the price of loosing the unit norm for all cases but $N=2$, where we can write $\tilde{\mathcal{J}}_{k} (\xi/2 ) = 2 \tilde{\mathcal{S}}_{k} ( \zeta)$ with $j=1/2$.
Note that this projection also allows us to derive an asymptotic expression for the $z$-component of the Stokes vector via Eq.(\ref{eq:AsympNPlanar}), and heuristically propose the rest,
\begin{eqnarray}
\lim_{\xi \rightarrow \infty} \tilde{\mathcal{J}}_{x}\left(\xi\right)&=& 0,\\
\lim_{\xi \rightarrow \infty} \tilde{\mathcal{J}}_{y}\left(\xi\right)&=& - \frac{1}{\gamma} ~~j,\\
\lim_{\xi \rightarrow \infty} \tilde{\mathcal{J}}_{z}\left(\xi\right)&=& - \frac{ \sqrt{\gamma^{2} -1}}{\gamma} ~~j, \\
\lim_{\xi \rightarrow \infty} \tilde{\mathcal{J}}\left(\xi\right)&=& j,
\end{eqnarray}
this was confirmed numerically over a random sample of initial states and gain to coupling ratios outside the $\mathcal{PT}$-symmetric regime, $j \in \left[1/2, 10 \right]$ and $\gamma \in \left[1,3\right]$.
Figure \ref{fig:Figure11} shows the propagation of the renormalized Stokes vector in a six-waveguide coupler, $j=5/2$, with parameters in the $\mathcal{PT}$-symmetric, $\gamma = 0.5$, Fig. \ref{fig:Figure11}(a), fully degenerate, $\gamma = 1$, Fig. \ref{fig:Figure11}(b), and broken symmetry, $\gamma = 1.5$, Fig. \ref{fig:Figure11}(c), regimes for light impinging the first waveguide of the coupler in black.
Also, the renormalized Stokes vector propation for an initial field amplitude distribution corresponding to the eigenstate of $J_{x}$ with eigenvalue $-j$ is plotted in red to show the asymptotic behavior outside the $\mathcal{PT}$-symmetric regime.

%As a final note on the $PT$-symmetric $N$-waveguide coupler with underlying %$SU(2)$ symmetry, we want to bring forward that an alternate approach %involving the complexification of $SU(2)$,
%\begin{equation}
% \{ \mathbbm{J}_{x}, \mathbbm{J}_{y}, \mathbbm{J}_{z}\} \rightarrow \{ %\mathbbm{J}_{x}, i \mathbbm{J}_{y}, i \mathbbm{J}_{z}\}\equiv %\{\mathbbm{K}_{z},\mathbbm{K}_{x},\mathbbm{K}_{y}\},
%\end{equation}
%can be followed in order to obtain a nonunitary finite representation of %$SO(2,1)$, an optical realization of the Lorentz group in $2D+1$.
%Dealing with such a nonstandard representation of $SO(2,1)$ in detail may %require a discussion on its own due to the peculiarities induced by using a %finite representation of a noncompact group while, from a practical point of %view, addressing the system in the diagonal basis of $J_{z}$ already provides %us with with a propagator feasible of physical interpretation %\cite{RodriguezLara2015p5682}.

\section{Non-Hermitian Ehrenfest theorem and generalized Stokes vector}

So far, we have studied propagation of classical light through a class of $\mathcal{PT}$-symmetric devices with underlying $SO(2,1)$ symmetry that includes the linear and nonlinear $\mathcal{PT}$-symmetric dimer.
We have seen that propagation on these linear and nonlinear devices is more involved than in the passive case, $\gamma=0$, but the linear algebra approach has helped us produce closed form propagators for the lineal case and the renormalized fields approach and its Stokes vector representation has allowed us to find stationary states and visualize propagation dynamics.
Here, we shall try to understand the underlying reasons for this more complex propagation behavior.

As we mentioned earlier, the propagation of light through tight-binding $N$-waveguide couplers can be modeled by mode coupling theory in a form similar to the Schr\"odinger equation, Eq. (\ref{eq:SchrodingerEq}).
Thus, finding a propagator, Eq. (\ref{eq:PropagatedField}), provides us with the information of the complex field amplitudes as they propagate through each waveguide.
In quantum mechanics, we can also ask about the propagation of the expectation value, which in the optical picture translates to the following,
\begin{eqnarray}
\mathcal{A}(z) = \langle \mathcal{E}(z) \vert  \mathbbm{A}(z) \vert \mathcal{E}(z) \rangle,
\end{eqnarray}
for the operator $\mathbbm{A}(z)$, which can vary with propagation.
This is exactly what has been done when studying the propagation of the Stokes vector defined as the expectation value of the $SU(2)$ generators, Eqs. (\ref{StockesVector}) and (\ref{NStockesVector}), with the peculiarity that we used renormalized field amplitudes instead of just the field amplitudes in the linear cases.
Ehrenfest theorem relates the variation with propagation of the mean value with the dynamics of the model,
\begin{eqnarray}
\frac{d}{dz} \mathcal{A}(z) = \left\langle \frac{d}{dz} \mathbbm{A}(z) \right\rangle,
\end{eqnarray}
in standard Hermitian quantum mechanics, Heisenberg equation provides the equation of motion for the operator but in our non-Hermitian model we need to go beyond this.
Any non-Hermitian operator $\mathbbm{A}(z)$ can be decomposed,
\begin{eqnarray}
\mathbbm{A}(z) = \mathbbm{A}_{H}(z) + \mathbbm{A}_{S}(z),
\end{eqnarray}
in an Hermitian, $\mathbbm{A}_{H}^{\dagger}(z) = \mathbbm{A}_{H}(z)$, and skew-Hermitian,  $\mathbbm{A}_{S}^{\dagger}(z) = -\mathbbm{A}_{S}(z)$, parts.
Thus, we can define a commutator between non-Hermitian operators,
\begin{eqnarray}
\left[  \mathbbm{A}(z), \mathbbm{B}(z) \right]_{NH} = \mathbbm{A}(z) \mathbbm{B}(z) - \mathbbm{B}^{\dagger}(z) \mathbbm{A}^{\dagger}(z),
\end{eqnarray}
such that we can write a Heisenberg-like equation of motion ruled by a non-Hermitian Hamiltonian,
\begin{eqnarray}
\frac{d}{dz} \mathbbm{A}(z) = i \left[  \mathbbm{A}(z), \mathbbm{H}(z) \right]_{NH} + \frac{\partial}{\partial z} \mathbbm{A}(z),
\end{eqnarray}
and recover a non-Hermitian generalization of Ehrenfest theorem for tight binding non-Hermitian waveguide couplers,
\begin{eqnarray}
\frac{d}{dz} \mathcal{A}(z) =  i \left\langle \left[  \mathbbm{A}(z), \mathbbm{H}(z) \right]_{NH} \right\rangle + \left\langle \frac{\partial}{\partial z} \mathbbm{A}(z) \right\rangle.
\end{eqnarray}
Let us consider as example the $N$-waveguide coupler of last section, Eq. (\ref{eq:HSU2}), where the effective mode-coupling matrix acting as Hamiltonian has as Hermitian part, $\mathbbm{H}_{H}=\mathbbm{J}_{x}$, and a skew-Hermitian part, $\mathbbm{H}_{S}= i \gamma \mathbbm{J}_{z}$.
Here the Ehrenfest theorem can be simplified to the expression,
\begin{eqnarray}
\partial_{\xi} \mathcal{A}(\xi) = i \left\langle \left[  \mathbbm{A}(\xi), \mathbbm{H}_{H} \right] \right\rangle +  i \left\langle \left\{  \mathbbm{A}(\xi), \mathbbm{H}_{S} \right\}  \right\rangle + \left\langle \partial_{\xi} \mathbbm{A}(\xi) \right\rangle,
\end{eqnarray}
where the standard commutator, $\left[ \mathbbm{A}, \mathbbm{B}\right] = \mathbbm{A} \mathbbm{B}  - \mathbbm{B}  \mathbbm{A}$, and anti-commutator, $\left\{ \mathbbm{A}, \mathbbm{B} \right\} = \mathbbm{A} \mathbbm{B}  + \mathbbm{B}  \mathbbm{A}$, have been used.
We can see that the anti-commutator term in this expression will be proportional to the gain to coupling strength ratio, $\gamma$, and this is the culprit behind the more complex behavior of our general class of $N$-waveguide lattices with underlying complexified $SU(2)$ symmetry.
Note that we can also use this result to derive the conserved quantities of the model by solving $\partial_{\xi} \mathcal{A}(\xi) = i \left\langle \left[  \mathbbm{A}(\xi), \mathbbm{H}_{H} \right] \right\rangle +  i \left\langle \left\{  \mathbbm{A}(\xi), \mathbbm{H}_{S} \right\}  \right\rangle + \left\langle \partial_{\xi} \mathbbm{A}(\xi) \right\rangle = 0$.

As a practical example, let us derive the equations of motion for the Stockes vector for the linear $\mathcal{PT}$-symmetric dimer, $\mathcal{S}_{k} = \langle \sigma_{k} \rangle$ with $k=0,x,y,z$, where $\sigma_{0}$ is the identity matrix, given by Eqs.(\ref{eq:SVx})--(\ref{eq:SV0}).
Here, the Hermitian and skew-Hermitian parts of the mode coupling matrix are
$\mathbbm{H}_{H}=\sigma_x$ and $\mathbbm{H}_{S}=i\gamma\sigma_z$.
Now, any linear Hermitian operator for this system can be written as the linear superposition of the matrices $\sigma_{k}$,
\begin{eqnarray}
\mathbbm{A}(\zeta)=\sum_{k=0,x,y,z} a_{k}^{(\mathbbm{A})}(\zeta) \sigma_\mu,
\end{eqnarray}
and the propagation of its expectation values, according to the non-Hermitian Ehrenfest theorem, are given by the following expression,
\begin{eqnarray}
\partial_\zeta \mathcal{A}(\zeta)&=& 2 \left[ a_{y}(\zeta) \mathcal{S}_{z}(\zeta) - a_{z}(\zeta) \mathcal{S}_{y}(\zeta) \right]  +   \sum_{k=0,x,y,z} \mathcal{S}_{k}(\zeta) \partial_\zeta a_{k}(\zeta) + \nonumber \\
&&- 2 \gamma \left[a_{0}(\zeta) \mathcal{S}_{z}(\zeta) + a_{z}(\zeta) \mathcal{S}_{0}(\zeta) \right]
\end{eqnarray}
Thus, noting that for the Stokes vectors the coefficients are constant, $a_{k}^{(\sigma_{l})}(\zeta)= \delta_{l,k}$, we can write the evolution for the components of the Stokes vector without field renormalization,
\begin{eqnarray}\label{eq:StockesEq}
  \partial_{\zeta} {\cal S}_{0}(\zeta) &=& - 2 \gamma \mathcal{S}_{z}(\zeta), \\
  \partial_{\zeta} {\cal S}_{x}(\zeta) &=& 0,  \\
  \partial_{\zeta} {\cal S}_{y}(\zeta) &=& 2 \mathcal{S}_{z}(\zeta),  \\
  \partial_{\zeta} {\cal S}_{z}(\zeta) &=& - 2 \left[ \mathcal{S}_{y}(\zeta) + 2 \gamma \mathcal{S}_{0}(\zeta) \right] ,
\end{eqnarray}
which are in complete agreement with what we obtain from the nonlinear $\mathcal{PT}$-symmetric dimer, Eqs. (\ref{eq:SVNL0})--(\ref{eq:SVNLz})  , if we kill the effective nonlinearity to coupling strength ratio, $\kappa = 0$.
Note that the total intensity, $\mathcal{S}_{0}(\zeta)$, is not conserved as expected from non-Hermitian dynamics.
Note that, in this case, the total intensity, $\mathcal{S}_{0}(\zeta)$, coincides with the norm of the Stokes vector, $\mathcal{S}(\zeta) =  \sqrt{ \mathcal{S}_{x}^{2}(\zeta) + \mathcal{S}_{y}^{2}(\zeta) + \mathcal{S}_{z}^{2}(\zeta) }$, and we can recover the renormalized Stokes vector dynamics, Eqs. (\ref{eq:RNSV0})--(\ref{eq:RNSVz}), if we define a renormalized Stokes vector, $\tilde{\mathcal{S}}_{k} = \mathcal{S}_{k}/\mathcal{S}_{0}$, and use the equations of motion found here.
If we were to find a constant of motion, $\mathbbm{S}_{c}(\zeta)$, then its components should satisfy,
\begin{eqnarray}
\partial_{\zeta} a_{0}^{(\mathbbm{S}_{c})}(\zeta) &=& 2 \gamma a_{z}^{\mathbbm{S}_{c}}(\zeta), \\
\partial_{\zeta} a_{x}^{(\mathbbm{S}_{c})}(\zeta) &=& 0, \\
\partial_{\zeta} a_{y}^{(\mathbbm{S}_{c})}(\zeta) &=& 2 a_{z}^{(\mathbbm{S}_{c})}(\zeta), \\
\partial_{\zeta} a_{z}^{(\mathbbm{S}_{c})}(\zeta) &=& 2 \left[ \gamma a_{0}^{(\mathbbm{S}_{c})}(\zeta) - a_{y}^{(\mathbbm{S}_{C})}(\zeta) \right].
\end{eqnarray}
A particular solution to this set of equation is $a_{k}^{(\mathbbm{S}_{c})}(\zeta) = \delta_{k,x}$ in agreement with Eq.(105).

For the general case of the planar $N$-waveguide coupler, the situation is far more complex as we are dealing with square matrices of dimension $N$.
In order to construct any given Hermitian operator of this dimension, we need a basis with  total of $N^{2}$ matrices, these are provided by the standard unitary group of degree $N$, $SU(N)$, plus the identity.
This way, we will work with a set of $N^{2}$ operators where the first four elements are the representation of $SU(2)$ in dimension $N$ plus the unity, $\mathbbm{J}_{k}$ with $k=0,x,y,z$, that form $U(N)$,
\begin{eqnarray}
\mathbbm{A} (\xi) = \sum_{k=0}^{N^{2}-1} a_{k}^{(\mathbbm{A})}(\xi) \mathbbm{J}_{k},
\end{eqnarray}
where we have just implicitly make the change $\mathbbm{J}_{x} = \mathbbm{J}_{1}$, $\mathbbm{J}_{y} = \mathbbm{J}_{2}$, $\mathbbm{J}_{z} = \mathbbm{J}_{3}$.
Thus, if we define a generalized Stokes vector for the planar $N$-waveguide coupler, it will have dimension $N^{2}$ and the zeroth component will be the total intensity in the system,
\begin{eqnarray}
\mathcal{J}_{0}(\xi) = \sum_{k=1}^{N} \vert \mathcal{E}_{k} (\xi) \vert^{2},
\end{eqnarray}
but in this case the zeroth component of the generalized Stokes vector, $\mathcal{J}_{0}(\zeta)$, is still the total intensity but does not coincide with the norm of the generalized Stokes vector, $\mathcal{J} = \sqrt{ \sum_{k} \vert \mathcal{J}_{k} \vert^{2} }$.
Thus, a graphical representation on the sphere will just be a projection of the propagation dynamics occurring on a $(N^{2}-1)$-dimensional hypersphere as mentioned before.
Note that for passive devices, $\gamma=0$, the propagation equations for the Stokes vectors do not involve any other functions and the propagation dynamics is restricted to the subgroup $SU(2)$ of $U(N)$, recovering the results of Ref. \cite{VillanuevaVergara2015p}

%%%%%%%%%%%%%%%%%%%%%%%%%%%%%%%%%%%%%%%%%%
\section{Quantum $PT$-symmetric dimer}

Let us turn our attention now to the propagation of nonclassical light.
In the quantum regime, it is possible to describe two-waveguide couplers with the following effective Hamiltonian \cite{Politi2008p646},
\begin{eqnarray}
\hat{H} = n_{1} \hat{a}_{1}^{\dagger} \hat{a}_{1} + n_{2}  \hat{a}_{2}^{\dagger} \hat{a}_{2}  + g \left( \hat{a}_{1}^{\dagger} \hat{a}_{2} + \hat{a}_{1} \hat{a}_{2}^{\dagger} \right),
\end{eqnarray}
where we have kept the notation for the effective refractive indices and evanescent coupling strength, $n_{j}$ with $j=1,2$ and g, in that order.
At this point, we can use Schwinger two-boson representation of $SU(2)$ \cite{Sattinger2013},
\begin{eqnarray}
J_{x} &=& \frac{1}{2} \left( \hat{a}_{1}^{\dagger} \hat{a}_{2} + \hat{a}_{1} \hat{a}_{2}^{\dagger} \right), \\
J_{y} &=& -\frac{i}{2} \left( \hat{a}_{1}^{\dagger} \hat{a}_{2} - \hat{a}_{1} \hat{a}_{2}^{\dagger} \right), \\
J_{z} &=& \frac{1}{2} \left( \hat{a}_{1}^{\dagger} \hat{a}_{1} - \hat{a}_{2}^{\dagger} \hat{a}_{2} \right),
\end{eqnarray}
to write an effective Hamiltonian with underlying $SU(2)$ symmetry,
\begin{eqnarray}
H = \omega J_{z} + J_{x}, \qquad \omega = \frac{1}{g} \left(n_{1} -n_{2} \right),
\end{eqnarray}
that answers to the effective Schr\"odinger equation,
\begin{equation}
-i \partial_{\xi} \vert \mathcal{E}(\xi) \rangle = H \vert \mathcal{E}(\xi) \rangle, \qquad \xi = 2g z.
\end{equation}
In the single photon regime,
\begin{eqnarray}
\vert \mathcal{E} (\xi) \rangle = \mathcal{E}_{1}(\xi) \vert 1,0 \rangle + \mathcal{E}_{2}(\xi) \vert 0,1 \rangle,
\end{eqnarray}
we recover the differential equation set describing the standard two-waveguide coupler,
\begin{eqnarray} \label{eq:StdDimer}
- i \partial_{\xi} \left( \begin{array}{c} \mathcal{E}_{1}(\xi) \\ \mathcal{E}_{2}(\xi) \end{array}\right) = \left( \begin{array}{cc}
\omega & 1 \\ 1 & -\omega  \end{array} \right) \left( \begin{array}{c} \mathcal{E}_{1}(\xi) \\ \mathcal{E}_{2}(\xi) \end{array}\right).
\end{eqnarray}
This approach suffices for the analysis of ideal dimers without gain nor loses, $\omega \in \mathbb{R}$, where the total photon number,
\begin{eqnarray}
\hat{n} = \hat{a}^{\dagger}_{1} \hat{a}_{1} + \hat{a}^{\dagger}_{2} \hat{a}_{2}
\end{eqnarray}
of the initial state, $\langle \hat{n} \rangle = \langle \mathcal{E}(0) \vert \hat{n} \vert \mathcal{E}(0) \rangle$, determines the dimension of the $SU(2)$ representation to be used,  $N = \langle \hat{n} \rangle + 1 = 2j +1$  , with the eigenbasis of $J_{z}$ given by the following,
\begin{eqnarray}
\vert j, m \rangle = \vert n_{1} , n_{2} \rangle, \qquad j = \frac{n_{1}+n_{2}}{2},~~ m=\frac{n_{1}-n_{2}}{2},
\end{eqnarray}
such that we can use the results proposed for classical waveguides couplers with underlying $SU(2)$ symmetry \cite{VillanuevaVergara2015p} to calculate relevant quantities like  mean photon number at each waveguide.

The inclusion of linear loses and gain is not a trivial matter and it is simpler to discuss in the Heisenberg picture \cite{Agarwal2012p031802},
\begin{eqnarray}
\frac{d}{dz} \hat{O}(z) = -i \left[ \hat{H}, \hat{O}(z) \right] + \partial_{z} \hat{O}(z),
\end{eqnarray}
where we have accounted for the change from time to distance propagation.
Let us go straight to the $\mathcal{PT}$-symmetric dimer, with identical real part of the refractive index and moving into a rotating frame,
\begin{eqnarray}
\left( \begin{array}{c} \hat{a}_{1}(z) \\ \hat{a}_{2}(z) \end{array}\right) = e^{  i n_{+} z } \left( \begin{array}{c} \hat{o}_{1}(z) \\ \hat{o}_{2}(z) \end{array}\right),
\end{eqnarray}
such that, again, we can define a scaled propagation, $\zeta = g z$, and include spontaneous processes arising from the quantum description of materials with linear loss or gain processes \cite{Agarwal2012p031802},
\begin{eqnarray}
&& \frac{d}{d\zeta} \left( \begin{array}{c} \hat{o}_{1}(\zeta) \\ \hat{o}_{2}(\zeta) \end{array}\right) = i ~ \mathbbm{H} \left( \begin{array}{c} \hat{o}_{1}(\zeta) \\ \hat{o}_{2}(\zeta) \end{array}\right) + \mathbbm{1} \left( \begin{array}{c} \hat{f}_{1}(\zeta) \\ \hat{f}_{2}(\zeta) \end{array}\right), \\
&& \mathbbm{H}= \left( \begin{array}{cc}
i \gamma & 1 \\ 1 & -i\gamma  \end{array} \right)
\end{eqnarray}
where the first term in the right hand side is related to propagation through the quantum two-waveguide coupler with linear loss and gain, $\gamma \in \mathbb{R}$, and the second term describe Gaussian random processes of emission and absorption, a result arising from the linear materials in an equivalent treatment to that used in the quantum description of the laser \cite{Scully2001},
\begin{eqnarray}
\langle \hat{f}_{1}^{\dagger}(\zeta) \hat{f}_{1}(\zeta^{\prime}) \rangle &=& 0, \qquad \quad \qquad  \langle \hat{f}_{1}(\zeta) \hat{f}_{1}^{\dagger}(\zeta^{\prime}) \rangle = 2\gamma~ \delta(\zeta-\zeta^{\prime}), \\
\langle \hat{f}_{2}^{\dagger}(\zeta) \hat{f}_{2}(\zeta^{\prime}) \rangle &=& 2\gamma~ \delta(\zeta-\zeta^{\prime}), \qquad ~ \langle \hat{f}_{2}(\zeta) \hat{f}_{2}^{\dagger}(\zeta^{\prime}) \rangle = 0.
\end{eqnarray}
The formal solution for this differential equation yields the propagation of the annihilation operators,
\begin{eqnarray}
\left( \begin{array}{c} \hat{a}_{1}(\zeta) \\ \hat{a}_{2}(\zeta) \end{array}\right) = e^{i \mathbbm{H} \zeta} \left( \begin{array}{c} \hat{a}_{1}(0) \\ \hat{a}_{2}(0) \end{array}\right) + \int_{0}^{\zeta} e^{i \mathbb{H} \left(\zeta - t\right)} \left( \begin{array}{c} \hat{f}_{1}(t) \\ \hat{f}_{2}(t) \end{array}\right)dt,
\end{eqnarray}
where we have obviated the common phase factor $e^{i n_{+} \zeta / g}$ that does not play any important role.
Note that we can use the propagator we already found for the classical dimer, $ \mathbbm{U} = e^{i \mathbb{H} \zeta}$ in Eq. (\ref{eq:Propagator}), for the first term in the right hand side.

\begin{figure}[h]
\centering
\includegraphics[width=\textwidth]{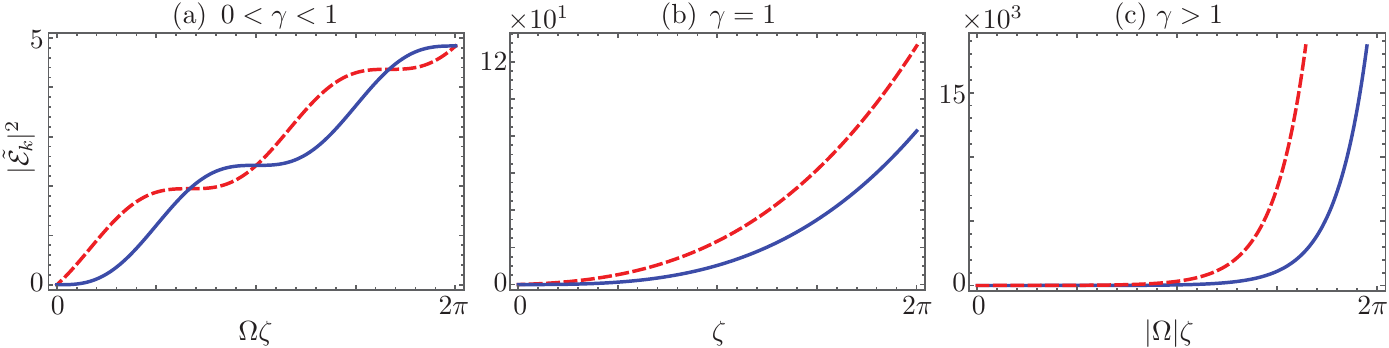}
\caption{Spontaneous generation of radiation in the waveguides with effective loss, $ S_{1}$ solid blue line, and gain, $S_{2}$ dashed red line, in the (a) $\mathcal{PT}$-symmetric regime, $\gamma=0.5$, (b) fully degenerate regime, $\gamma=1$, and (c) broken symmetry regime, $\gamma=1.5$, for quantum vacuum fields in both waveguides. }
\label{fig:Figure12}
\end{figure}

Now, in order to realize the effect of processes induced by the linear materials, let us focus on spontaneous generation in the absence of fields in both waveguides.
In the classical case, there will be no light at all propagating through the waveguides but, in the quantum case, even with an initial vacuum state we can calculate the spontaneous generation at each waveguide \cite{Agarwal2012p031802},
\begin{eqnarray}
S_{1} &=& 2 \gamma \int_{0}^{\zeta} \left\vert \left[e^{i \mathbbm{H} t}\right]_{12} \right\vert^{2} dt, \\
S_{2} &=& 2 \gamma \int_{0}^{\zeta} \left\vert \left[e^{i \mathbbm{H} t}\right]_{22} \right\vert^{2} dt.
\end{eqnarray}
It shows in the symmetric regime, $\gamma <1$,
\begin{eqnarray}
S_{1} &=& \frac{1}{4 \Omega^{3}} \left[ 2 \Omega \zeta - \sin \left(2 \Omega \zeta \right) \right], \\
S_{2} &=& \frac{1}{4 \Omega^{3}} \left[ 2 \Omega \zeta + \left(1- 2\gamma^{2} \right) \sin\left(2 \Omega \zeta \right) + 4 \gamma \Omega \sin^{2} \left(2 \Omega \zeta \right) \right],
\end{eqnarray}
a linear increase with a periodic modulation, Fig. \ref{fig:Figure12}(a).
In the fully degenerate case, $\gamma = 1$, the spontaneous generation,
\begin{eqnarray}
S_{1} &=& \frac{1}{3} \zeta^{3}, \\
S_{2} &=& \frac{\zeta}{3} \left(3 + 3 \zeta + \zeta^{2}\right),
\end{eqnarray}
follows a cubic polynomial, Fig. \ref{fig:Figure12}(b), and in the broken symmetry regime, $\gamma > 1$,
\begin{eqnarray}
S_{1} &=& \frac{1}{4 \vert\Omega\vert^{3}} \left[ -2 \vert \Omega \vert \zeta + \sinh \left(2 \vert \Omega \vert \zeta \right) \right], \\
S_{2} &=& \frac{1}{4 \vert \Omega \vert^{3}} \left[- 2 \vert \Omega \vert \zeta - \left(1- 2\gamma^{2} \right) \sinh \left(2 \vert \Omega \vert \zeta \right) + 4 \gamma \vert \Omega \vert \sinh^{2} \left(2 \vert \Omega \vert \zeta \right) \right],
\end{eqnarray}
it shows exponential amplification, Fig. \ref{fig:Figure12}(c).
Further discussion regarding the effect of spontaneous processes on the propagation of diverse nonclassical fields through a linear $\mathcal{PT}$-symmetric dimer can be found in Ref. \cite{Agarwal2012p031802}.

%%%%%%%%%%%%%%%%%%%%%%%%%%%%%%%%%%%%%%%%%%
%\section{Materials and Methods}

%%%%%%%%%%%%%%%%%%%%%%%%%%%%%%%%%%%%%%%%%%
\section{Conclusions}

We have presented a review of the $\mathcal{PT}$-symmetric dimer in its linear, nonlinear and quantum versions and show that it belongs to a symmetry class with underlying $SO(2,1)$ symmetry, realized as a complexification of the  $SU(2)$ group, that allows the description of $N$-waveguide couplers.
We have aimed to present a coherent narrative of the different approaches to the optical $\mathcal{PT}$-symmetric dimer and relate them to the underlying symmetry of the model.
In doing this, we introduce the idea of using a non-Hermitian version of Ehrenfest theorem to approach the propagation dynamics of waveguide couplers described by non-Hermitian mode coupling matrices.

The field is young and there still exist fundamental open questions on the subject such as the analytic determination of critical effective nonlinearity to coupling ratios for the Kerr nonlinear $\mathcal{PT}$-symmetric dimer; the need of a deeper understanding of the non-unitary finite dimensional representations of $SO(2,1)$, realized without resorting to the complexified $SU(2)$ representations; the generalization to propagation dependent photonic systems together with its possible applications, just to mention a few that we hope to address in future work.

%%%%%%%%%%%%%%%%%%%%%%%%%%%%%%%%%%%%%%%%%%
\section*{Acknowledgments}
DHM acknowledges financial support from CONACYT $\#$294921 PhD grant. SLA acknowledges financial support from CONACYT $\#$243284 grant.

%%%%%%%%%%%%%%%%%%%%%%%%%%%%%%%%%%%%%%%%%%
\section*{References}

%\bibliographystyle{unsrt}
%\bibliography{references}

\end{document}